\documentclass[]{aa}
\usepackage{graphics,latexsym,epsfig,graphicx,amssymb}     

\renewcommand{\d}[0]{{\rm d}}

\newcommand{\ave}[1]{\langle #1 \rangle}

\newcommand{\Ref}[1]{(\ref{#1})}

\newcommand{\Vector}[1]{{\vec{#1}}}
\newcommand{\Matrix}[1]{{\tens{#1}}}
\newcommand{\pprime}[0]{{\prime\prime}}

\newcommand{\stressa}[1]{{#1}}

\begin{document}

\title{GaBoDS: The Garching-Bonn Deep Survey\thanks{Based on
    observations made with ESO Telescopes at the La Silla Observatory}}
\subtitle{VII. Probing galaxy bias using weak gravitational lensing}

\author{P. Simon$^1$, M. Hetterscheidt$^1$, M. Schirmer$^2$,
  T. Erben$^1$, P. Schneider$^1$,\\ C. Wolf$^3$, and K. Meisenheimer$^4$}

\institute{%
  $^1$Universit\"at Bonn, Argelander-Institut f\"ur
  Astronomie\thanks{Founded by merging of the Institut f\"ur
    Astrophysik und Extraterrestrische Forschung, the Sternwarte, and
    the Radioastronomisches Institut der Universit\"at Bonn.}
  Auf dem H\"ugel 71, 53121 Bonn, Germany,\\
  \email{psimon@astro.uni-bonn.de},\\
  $^2$ Isaac Newton Group of Telescopes, Santa Cruz de La Palma, Spain\\
  $^3$  University of Oxford, Denys Department of Physics, Wilkinson
  Building, Keble Road, Oxford OX1 3RH, U.K.\\
  $^4$ Max-Planck-Institut f\"ur Astronomie, K\"onigsstuhl 17, 69117
  Heidelberg, Germany}

\date{} \authorrunning{Simon et al.} \titlerunning{Galaxy bias in
  GaBoDS}


  \abstract{}{%
    An interesting question of contemporary cosmology concerns the
    relation between the spatial distribution of galaxies and dark
    matter, which is thought to be the driving force behind the
    structure formation in the Universe.  In this paper, we measure
    this relation, parameterised by the linear stochastic bias
    parameters, for a range of spatial scales using the data of the
    Garching-Bonn Deep Survey (GaBoDS).}{%
    The weak gravitational lensing effect is used to infer matter
    density fluctuations within the field-of-view of the survey
    fields. This information is employed for a statistical comparison
    of the galaxy distribution to the total matter distribution.  The
    result of this comparison is expressed by means of the linear bias
    factor $b$, the ratio of density fluctuations, and the correlation
    factor $r$ between density fluctuations.  The total galaxy sample
    is divided into three sub-samples using $R$-band magnitudes and
    the weak lensing analysis is applied separately for each
    sub-sample.  Together with the photometric redshifts from the
    related COMBO-17 survey we estimate the typical mean redshifts of
    these samples with \mbox{$\bar{z}=0.35, 0.47, 0.61$},
    respectively.}{%
    Using a flat $\Lambda\rm CDM$ model with \mbox{$\Omega_{\rm
        m}=0.3, \Omega_\Lambda=0.7$} as fiducial cosmology, we obtain
    values for the galaxy bias on scales between
    \mbox{$1^\prime\le\theta_{\rm ap}\le20^\prime$}. At $10^\prime$,
    the median redshifts of the samples correspond roughly to a
    typical comoving scale of \mbox{$3,5,7~h^{-1}\rm Mpc$} with
    \mbox{$h=0.7$}, respectively. We find evidence for a
    scale-dependence of $b$.  Averaging the measurements of the bias
    over the range \mbox{$2^\prime\le \theta_{\rm ap}\le 19^\prime$}
    yields $\bar{b}=0.81\pm0.11, 0.79\pm0.11, 0.81\pm0.11$
    ($1\sigma$), respectively.  Galaxies are thus less clustered than
    the total matter on that particular range of scales (anti-biased).
    As for the correlation factor $r$ we see no scale-dependence
    within the statistical uncertainties; the average over the same
    range is $\bar{r}=0.61\pm0.16, 0.64\pm0.18, 0.58\pm0.19$
    ($1\sigma$), respectively.  This implies a possible decorrelation
    between galaxy and dark matter distribution. An evolution of
    galaxy bias with redshift is not found, the upper limits are:
    $\Delta b\lesssim0.2$ and $\Delta r\lesssim 0.4(1\sigma)$.}{}

\maketitle

\section{Introduction}

In comparison to the total mass in the Universe, galaxies take --
considering their mass -- only a minor part in the big picture of
structure formation.  \stressa{They formed from the baryonic
  component, embedded in the fluctuations of the dark matter density
  field, whose total mean density is very much lower than that of dark
  matter.}  Due to their relatively easy observability, it would be
very convenient if galaxies were perfect tracers -- unbiased tracers
-- of the total mass distribution; all statistical properties of the
mass structure could then be derived from galaxy catalogues.

Indeed, it is rather unlikely that galaxies are unbiased tracers,
because the laws determining the galaxy distribution are very complex
and highly non-linear.  The primordial gas from which they form
requires special conditions to be able to cool and fragment into
galaxies (White~\&~Frenk 1991; White~\&~Rees 1978). Due to shock
heating of the baryons and energy feedback between galaxies and the
intergalactic medium, the properties of the gas feeding galaxy
formation gradually changed with time.  Furthermore, galaxies interact
with each other or with the baryonic intergalactic medium, merge or
get accreted into more massive galaxies (Lacey~\&~Cole 1993).  These
mechanisms probably produced the large diversity in galaxy masses,
colours, morphologies and chemistry we observe today.  Based on our
current knowledge it would be very surprising if this complexity would
eventually result in a simple, linear, one-to-one relationship between
the galaxy density and total matter density, making galaxies unbiased
tracers.  But by studying this dark matter-galaxy relationship we can
learn more about galaxies.

Observing the relation between the invisible dark matter field and the
galaxies is a particularly tough problem.  However, with gravitational
lensing at hand, we now have a technique to directly unravel this
relationship. The importance of ``cosmic shear'' as a tool for
cosmology was proposed in the early 1990s by Blandford et al. (1991),
Miralda-Escud\'e (1991) and Kaiser (1992). Since these pioneering days
of gravitational lensing the techniques and surveys have been refined
to make valuable contributions to the ongoing research on structure
formation on cosmological scales. In particular, the investigation of
the relation between galaxy and dark matter distribution using the
weak gravitational lensing effect has become almost standard (Seljak
et al.  2005; Kleinheinrich et al.  2005; Sheldon et al. 2004;
Hoekstra et al.  2003; Guzik~\&~Seljak 2001; McKay et al. 2001; Fisher
et al. 2000; Hudson et al. 1998; Brainerd et al. 1996). This paper
will focus on the lensing technique as well.

\subsection{Quantifying galaxy bias}

From the point of view of statistics, quantifying galaxy bias leads to
the question how one can parametrise differences in the statistical
properties -- not the obvious differences between two particular
realisations -- of two random fields. Both the distribution of
galaxies and the distribution of dark matter are thought to be
realisations of statistically homogeneous and isotropic random fields.
Commonly, one uses a parametric way to describe the biasing between
two (random) density fields, for instance galaxy and matter
distribution or the distributions of two different galaxy populations.
Biasing between two density fields, say $\rho_{\rm g}$ and $\rho_{\rm
  m}$, can in general be quantified using the joint probability
distribution function (PDF) $P\left(\delta_{\rm g},\delta_{\rm
    m}\right)$ of the density contrasts (density fluctuations)
\begin{equation}
  \delta_{\rm g}\equiv\frac{\rho_{\rm g}}{\ave{\rho_{\rm g}}}-1~~;~~
  \delta_{\rm m}\equiv\frac{\rho_{\rm m}}{\ave{\rho_{\rm m}}}-1
\end{equation}
at the same point in the density fields at some redshift (local
Eulerian bias). The probability of finding density contrasts of
$\delta_{\rm g}$ and $\delta_{\rm m}$ within an interval of
$\d\delta_{\rm g}$ and $\d\delta_{\rm m}$, respectively, equals
$P(\delta_{\rm g},\delta_{\rm m})\d\delta_{\rm m}\d\delta_{\rm g}$.
The density contrasts are smoothed to a certain scale, $R$, before
investigating their PDF.  Often the special type of smoothing kernel
is set by the method that is used to determine the bias.  To lowest,
second order (first order moments vanish by definition,
$\ave{\delta_{\rm g}}=\ave{\delta_{\rm m}}=0$) the only relevant
parameters for biasing are the scale-dependent parameters of the
``linear stochastic bias''
\begin{equation}\label{biascoeff}
  b(R)=\sqrt{\frac{\ave{\delta_{\rm g}^2}}{\ave{\delta_{\rm m}^2}}}
  ~~;~~
  r(R)=\frac{\ave{\delta_{\rm g}\delta_{\rm m}}}{\sqrt{\ave{\delta_{\rm
          g}^2}\ave{\delta_{\rm m}^2}}}
  \; ,
\end{equation}
which differ from unity in the case of two biased fields. These two
parameters provide a complete description for the bias between
$\delta_{\rm g}$ and $\delta_{\rm m}$ for Gaussian fields only and
therefore are a full description only on large smoothing scales where
linear perturbation theory applies (on ``linear scales''). The linear
bias factor $b$ is a measure for the difference in clustering
strength. The correlation coefficient $r$, on the other hand, measures
partly the stochasticity in the relation between the density
contrasts, for instance how well minima and maxima of the density
fields coincide. Only partly, because the correlation coefficient is
also sensitive to the non-linearity in this relation, which becomes
relevant on smaller smoothing scales where the fields are usually no
longer Gaussian. In fact, they cannot be Gaussian since
\stressa{$\delta\ge-1$} by definition. Despite this degeneracy, the
linear stochastic bias parameters are clearly defined on non-linear
scales; it is just their interpretation which is no longer
straightforward. To disentangle a non-linear but deterministic
relation between two density contrasts from a stochastic one requires
higher-order statistics, like for example in the context of the
``non-linear stochastic bias'' scheme (Dekel~\&~Lahav 1999).

\subsection{Galaxy bias in observations}

Observationally, galaxy bias can be derived from the one-dimensional
PDF of galaxies, $P(\delta_{\rm g})$, (Marinoni et al. 2005; Sigad et
al.  2000), redshift space-distortions (Pen 1998; Kaiser 1987), weak
gravitational lensing (Seljak et al. 2005; Sheldon et al. 2004; Pen et
al. 2003; Hoekstra et al.  2002; Wilson et al. 2001; van Waerbeke
1998; Schneider 1998) and counts-in-cells statistics (Conway et al.
2005; Tegmark~\&~Bromley 1999; Efstathiou et al. 1990). Additionally,
the large-scale flow of galaxies can be used to make a POTENT
reconstruction of the total mass field on large scales which can be
compared to the galaxy distribution (Sigad et al. 1998; Dekel et al.
1993). Gravitational lensing (Schneider et al. 2006; van
Waerbeke~\&~Mellier 2003; Bartelmann~\&~Schneider 2001) provides a
promising new method in this respect because it allows for the first
time to map the total matter content (mainly dark matter) independent
from the galaxy distribution. The work of this paper is based on this
technique.

A brief overview of the current status of the observational results is
given in the following. Note that the given conclusions to some extent
depend on the assumed cosmological model. We quote only the
conclusions for the concordance $\Lambda\rm CDM$ model (cf.  Tegmark
et al. 2004). In the local universe, $L_\star$ galaxies are almost
unbiased tracers on linear scales of about $8h^{-1}\rm Mpc$ and larger
(Seljak et al. 2005; Verde et al.  2002; Lahav et al. 2002; Loveday et
al.  1996).  However, this is probably not true on smaller scales. A
comparison of the theoretical dark matter clustering -- which is
constrained by the cosmic microwave background anisotropies,
gravitational lensing and the Lyman-$\alpha$ forest (Tegmark et al.
2004) -- and the observable galaxy clustering suggests that on smaller
scales $\sim1h^{-1}\rm Mpc$ galaxies are less clustered than the dark
matter (``anti-biased'') becoming positively biased, $b>1$, on even
smaller scales below $\sim0.1h^{-1}\rm Mpc$. Hoekstra et al.  (2001,
2002) use in their work on the VIRMOS-DESCART survey (van Waerbeke et
al.  2001) and RCS (Gladders~\&~Yee 2001) weak gravitational lensing
to measure the linear stochastic bias for galaxies with a median
redshift of $z_{\rm m}=0.35$, covering a range from $0.1h^{-1}{\rm
  Mpc}$ to $6.3h^{-1}{\rm Mpc}$.  They claim to have observed such a
dip in the linear bias factor.  Also based on gravitational lensing
there is evidence that the ratio $b/r$ stays close to unity from
sub-megaparsec scales up to \mbox{$\sim8h^{-1}\rm Mpc$} (Sheldon et
al.  2004; Hoekstra et al.  2002; Guzik~\&~Seljak 2001; Fisher et al.
2000), thus from non-linear to linear scales. The analysis of the
bispectrum of the galaxy clustering in the 2dFGRS (Colless et al.
2001) led Verde et al. (2002) to the conclusion that on scales
between $5h^{-1}\rm Mpc$ and $30h^{-1}\rm Mpc$ the biasing relation
between dark matter and galaxies is essentially linear (see also Lahav
et al.  2002). The same conclusion was drawn several years earlier by
Gazta\'{n}aga \& Frieman (1994) based on the APM survey (Maddox et al.
1990).  However, recently the work of Kayo et al.  (2004) has
questioned a strict linear relation on scales
\mbox{$\lesssim10h^{-1}\rm Mpc$} by studying the three-point
correlation of galaxy clustering as a function of morphology, colour
and luminosity, this time in the SDSS (York et al. 2000).

Subdividing the galaxies into various subsets gives a more detailed
picture of galaxy biasing. At low redshift, clustering is a function
of morphological type, spectral type, colour and luminosity (e.g.
Madgwick et al. 2003;  Zehavi et al.  2002; Norberg et al.  2001;
Benoist et al.  1996; Tucker et al. 1997; Loveday et al.  1995;
Davis~\&~Geller 1976).  Late-type, blue, spiral or star forming
galaxies are less clustered than early-type, elliptical or red
galaxies with a relative linear bias factor of about \mbox{$b_{\rm
    red}/b_{\rm blue}\approx1.4$} on scales of roughly $8\,h^{-1}\rm
Mpc$ (e.g.  Wild et al.  2005; Conway et al.  2005; Zehavi et al.
2002; Norberg et al. 2002; Baker et al. 1998).  On large scales, the
relative biasing between red and blue galaxies does not seem to be
well described by a simple linear biasing function which according to
Wild et al. (2005) (see also Conway et al.  2005) is ruled out with
high significance using counts-in-cells statistics in redshift space.
Wild et al. observe a scale-dependent non-linear bias between red and
blue galaxies with a dominant stochasticity component for typical
physical scales of about $7h^{-1}\rm Mpc$ up to $31h^{-1}\rm Mpc$.
Relative bias between red and blue galaxies is therefore both
non-linear and stochastic. It is therefore also expected that at least
for some galaxy populations the bias with respect to the dark matter
distribution is non-linear and stochastic as suggested by simulations
(Yoshikawa et al. 2001). This, however, has not been measured directly
so far.  Observational evidence for the relation between red and blue
galaxies being non-deterministic was already given some years ago by the
work of Tegmark~\&~Bromley (1999), which was based on the Las Campas
Redshift Survey (Shectman et al.  1996), and by Blanton (2000).
Moreover, galaxy bias seems to be a function of redshift (Marinoni et
al.  2005; Magliocchetti et al.  2000) which is expected both from
simulations (Weinberg et al. 2004 and references therein) and
analytical models (Tegmark~\&~Peebles 1998; Mo~\&~White 1996; Fry
1996).

In this paper, we apply the method of Hoekstra et al. (2002; Sect.
\ref{methodsection}) to the Garching-Bonn Deep Survey (Sect.
\ref{gabodssection}) to obtain the linear stochastic bias coefficients
$b$ and $r$ of three galaxy subsets binned by their apparent $R$-band
magnitude. The selection of the galaxy samples is outlined in Sect.
\ref{gabodssection}.  After presenting our results in Sect.
\ref{resultsection} we close with a discussion and conclusions in
Sect.  \ref{finalsection}. We will start with a brief introduction to
the formalism of weak gravitational lensing and the aperture
statistics here employed.

Unless otherwise stated we use a $\Lambda$CDM model with
\mbox{$\Omega_{\rm m}=0.3$}, \mbox{$\Omega_{\Lambda}=0.7$} and
\mbox{$H_{0}=h\,100\,{\rm km}\,{\rm s}^{-1}\,{\rm Mpc}^{-1}$} with
\mbox{$h=0.7$}.  A scale-invariant (\mbox{$n=1$}, Harrison-Zel'dovich)
spectrum of primordial fluctuations is assumed.  As transfer function,
encoding the physical properties of the cold dark matter fluid, we use
Bardeen et al. (1986).

\section{Formalism}
\label{formalismsection}

\subsection{Weak gravitational lensing}

Weak gravitational lensing uses the shapes of distant galaxies -- the
source galaxies or, as we will also call them, background galaxies --
to infer the distribution of the total matter.  This is based on the
fact that light is deflected by density fluctuations so that the tidal
gravitational field of the matter density inhomogeneities along the
line-of-side towards a galaxy changes the shape of its image.  We
consider only cases in which the light rays emitted from a source
galaxy traverse only regions in space with relatively small
perturbations in the gravitational field (weak lensing regime); this
holds for almost every galaxy.

\paragraph{Theory.} In the weak lensing regime the differential
distortion effect of the tidal gravitational field is well described
by a two-dimensional linear mapping over the whole apparent size of
one galaxy. The relevant components of the linear transformation
$\Matrix{A}$ are the convergence $\kappa$, which magnifies or
demagnifies the size of a galaxy, and the shear,
$\gamma\equiv\gamma_1+{\rm i}\gamma_2$, which stretches the image of a
galaxy along some direction:
\begin{equation}
  \Matrix{A}(\Vector{\theta})=
  \left(
    \begin{array}{cc}
      1-\kappa-\gamma_1&-\gamma_2\\
      -\gamma_2&1-\kappa+\gamma_1
    \end{array}
  \right) \;,
\end{equation}
where $\Vector{\theta}$ is an angular position on the sky. Note that
for convenience 2D vectors are written as complex numbers, as for
instance $\Vector{\theta}=\theta_1+{\rm i}\theta_2$.

According to the theory of weak gravitational lensing, convergence and
shear are, to lowest order, a projection of the three-dimensional
density contrast $\delta_{\rm m}$ of the matter in the Universe via
\begin{eqnarray}\label{convergence}
  \kappa(\Vector{\theta})&=&
  \int_0^{w_{\rm h}} {\rm d} w\,
  \overline{W}\left(w\right)
  \delta_{\rm m}(f_{\rm K}\left(w\right)\Vector{\theta},w)\;,
  \\ \label{shear}
  \gamma(\Vector{\theta})&=&
  -\frac{1}{\pi}\int {\rm d^2}\Vector{\theta}^\prime\,
  \kappa(\Vector{\theta}^\prime-\Vector{\theta})
  \frac{1}{(\theta_1^\prime-{\rm i}\theta_2^\prime)^2}
  \; ,
\end{eqnarray}
where
\begin{equation}
  \overline{W}\left(w\right)=
  \frac{3\Omega_{\rm m}H_0^2f_{\rm K}\left(w\right)}{2c^2a\left(w\right)}
  \int_w^{w_{\rm h}}\!\!\!\!{\rm d}w^\prime\,p_{\rm b}\left(w^\prime\right)
  \frac{f_{\rm K}\left(w^\prime-w\right)}{f_{\rm
      K}\left(w^\prime\right)}
\end{equation}
is a weight given to the density contrast at a comoving radial
distance $w$ from the observer. The functions $a(w)$ and $f_{\rm
  K}(w)$ are the cosmic scale factor and the comoving angular
distance, respectively, at comoving distance $w$. The variable $w_{\rm
  h}$ is the comoving horizon size. By $p_{\rm b}(w)$ we denote the
distribution of the background sources in comoving distance. The shear
$\gamma$ as a function of the direction on the sky $\Vector{\theta}$
is called the cosmic shear map. According to Eq. \Ref{shear} it is a
convolution of the convergence $\kappa$, thus the projected matter
density contrast, with some kernel.

In the formalism presented here, we assume that we always observe only
small patches of the celestial sphere; small enough to approximate the
topology of the patch by a tangential, Cartesian plane (flat-sky
approximation). This is a very good approximation for the survey
fields considered here, which are smaller than $1\,\rm deg^2$.
Furthermore, the 3D coordinate system, such as for $\delta_{\rm m}$ in
Eq.  \Ref{convergence}, is chosen such that $w$ is a comoving distance
along some fixed reference line-of-sight and $f_{\rm
  K}(w)\Vector{\theta}$ a 2D-vector perpendicular to the reference
line-of-sight; $w$ is also used as look-back time parameter to account
for the fact that $\delta_{\rm m}$ is a function of time.

\paragraph{Connection to the real world.} To quantify the shape of a
galaxy one defines the complex ellipticity, \mbox{$\epsilon=\epsilon_1+{\rm
  i}\epsilon_2$}, which is related to the quadrupole moments of the
light distribution in the galaxy image (e.g. Schneider et al. 2006).
It transforms under rotations according to
\mbox{$\epsilon^\prime=\exp{(-2{\rm i}\psi)}\,\epsilon$} where $\psi$
is the rotation angle.

Seitz~\&~Schneider (1997), for example, showed that under the action
of the linear transformation $\Matrix{A}$ the \emph{intrinsic
  ellipticity}, i.e.  the unlensed galaxy ellipticity, $\epsilon_{\rm
  s}$, of a source galaxy is transformed into the image ellipticity,
$\epsilon$, according to
\begin{eqnarray}
 \epsilon\approx\epsilon_{\rm s}+\gamma
 \; ,
\end{eqnarray}
if $\gamma\ll 1$ which is fulfilled in the weak lensing regime.  The
practical importance of this equation stems from the fundamental
assumption that galaxies have intrinsically no preferred direction and
are therefore randomly oriented, i.e. $\ave{\epsilon_{\rm s}}=0$.
This makes the observed ellipticities of galaxies,
$\epsilon(\Vector{\theta})$, unbiased estimators of the cosmic shear
field in the direction $\Vector{\theta}$:
\begin{equation}
  \ave{\epsilon(\Vector{\theta})}=\gamma\left(\Vector{\theta}\right)
  \; . 
\end{equation} 
In this picture, the shear is a function of the convergence, and the
convergence is related to the projected $\delta_{\rm m}$.
Consequently, it should be possible to make a reconstruction of the
total matter distribution based on the observed ellipticities of
source galaxies, or a reconstruction of structural parameters such as
spatial correlation functions.

Of course, galaxies are in general not intrinsically round objects,
\mbox{$\epsilon_{\rm s}\ne 0$}; the ellipticities of single galaxies
have typically \mbox{$\left\ave{|\epsilon_{\rm s}\right|}\approx 0.3$}
\stressa{with a comparable scatter of
  \mbox{$\sigma(|\epsilon_s|)\approx0.3$}}.  This makes them in fact
very noisy estimators of the shear, considering that the shear signal
induced by gravitational lensing is typically about one percent of
this value. Therefore, the average over many galaxy ellipticities is
required in weak lensing applications.

\subsection{Aperture statistics}

\paragraph{Aperture mass.} For our application here, which aims at the
linear stochastic bias parameters on some smoothing scale, we are not
interested in the convergence field itself but in the convergence
field smoothed to some typical scale using an aperture filter $u$.  A
convenient quantity for this purpose is the aperture mass $M_{\rm ap}$
(Schneider et al. 1998) defined as
\begin{equation}
  M_{\rm ap}\left(\theta_{\rm ap},\Vector{\theta}\right)\equiv
      \frac{1}{\theta^2_{\rm ap}}\int
      {\rm d}^2\Vector{\theta}^\prime~
      u\left(\frac{\left|\Vector{\theta}^\prime-\Vector{\theta}\right|}{\theta_{\rm
      ap}}\right)\kappa\left(\Vector{\theta}^\prime\right) \; .
\end{equation} 
The variable $\theta_{\rm ap}$ is the aperture radius setting the
filter scale.  It has been shown that if the filter $u$ is
compensated, $\int {\rm d}x~x~u\left(x\right)=0$, then the aperture
mass can be determined from the shear field itself (Schneider 1996,
1998)
\begin{eqnarray}
\label{aperturemassdef}
  M_{{\rm ap}/\times}\left(\theta_{\rm ap},\Vector{\theta}\right)\!\!&=&\!\!
  \frac{1}{\theta_{\rm ap}^2}\int {\rm d}^2\Vector{\theta}^\prime~
  q\left(\frac{\left|\Vector{\theta}-\Vector{\theta}^\prime\right|}
    {\theta_{\rm ap}}\right)\!\gamma_{{\rm
    t}/\times}\left(\Vector{\theta}^\prime\right)\;, \\ 
  q\left(x\right)&\equiv& \frac{2}{x^2}\int_0^x {\rm d}s~s~ 
  u\left(s\right)-u\left(x\right) \; ,
\end{eqnarray}
where 
\begin{eqnarray}
  \gamma_{\rm t}\left(\Vector{\theta}^\prime\right)
  &\equiv&-{\rm Re}\left(\gamma(\Vector{\theta}^\prime){\rm
    e}^{-2{\rm i}\phi}\right)\;,\\
\gamma_\times\left(\Vector{\theta}^\prime\right)
&\equiv&-{\rm Im}\left(\gamma(\Vector{\theta}^\prime){\rm
    e}^{-2{\rm i}\phi}\right)
\end{eqnarray}
denote the tangential and cross shear component with respect to the
aperture centre $\Vector{\theta}$, respectively.  The variable $\phi$
is used as definition for the argument of the complex number
$\Vector{\vartheta}=\Vector{\theta}^\prime-\Vector{\theta}$, i.e.
$\vartheta_1+{\rm i}\vartheta_2=|\Vector{\vartheta}|e^{{\rm i}\phi}$.

In this definition, the actual aperture mass or so-called
\emph{E-mode} of the aperture mass is obtained by using the tangential
shear (with respect to the aperture centre), $\gamma_{\rm t}$, while
choosing the cross shear, $\gamma_\times$, gives the \emph{B-mode},
$M_\times$, of the aperture mass. As the shear originates from a
single scalar field, $\kappa$, the two shear components are related to
each other (cf. Schneider et al. 2002).  Therefore, not all
conceivable shear field configurations are produced by gravitational
lensing. The allowed configurations of $\gamma$ are called E-modes,
while the other independent configurations are called B-modes. For
that reason, a signature of B-modes is used in this paper as an
indicator for systematics in the data reduction, especially the
point-spread function (PSF) correction, which has to be performed to
compensate the instrumental and atmospheric influence on the galaxy
image. Note, however, that on scales smaller than about a few arcmin a
non-zero B-mode can be produced by intrinsic alignments of the source
galaxies (e.g. Heymans et al. 2004; Hirata et al.  2004) or spatial
clustering of the source galaxies (Schneider et al.  2002).

Since the ellipticity of a galaxy at $\Vector{\theta}$ is, in the weak
lensing regime, an unbiased estimator of the shear $\gamma$, one could
construct an estimator for the aperture mass that can be directly
applied to a galaxy catalogue in order to obtain a $M_{\rm ap}$-map
for some survey field (Hoekstra et al. 2001). We are, however,
interested in the relation between matter and galaxy distribution in a
statistical sense. As we will see soon, for this purpose it is not
even necessary to make an actual map -- even though this could be a
possible strategy. Before we discuss aperture statistics we introduce
a quantity to analyse the spatial distribution of galaxies.

\paragraph{Aperture number count.} In a similar fashion to the aperture
mass, we can define (Schneider 1998; van Waerbeke 1998) the aperture
number count $N(\Vector{\theta},\theta_{\rm ap})$ which measures the
fluctuations of the galaxy number density with the same filter $u$ as
$M_{\rm ap}$ for the convergence field:
\begin{eqnarray}\label{aperturecountdef}
  N(\Vector{\theta},\theta_{\rm ap})&\equiv&
  \frac{1}{\theta^2_{\rm ap}} \int {\rm d}^2\theta^\prime~
  u\left(\frac{|\Vector{\theta}-\Vector{\theta}^\prime|}
    {\theta_{\rm ap}}\right)\delta n(\Vector{\theta}^\prime)\\
  &=&\frac{1}{\bar{n}\theta^2_{\rm ap}}
  \int {\rm d}^2\theta^\prime~
  u\left(\frac{|\Vector{\theta}-\Vector{\theta}^\prime|}
    {\theta_{\rm ap}}\right) n(\Vector{\theta}^\prime)
  \; ,
\end{eqnarray}
where $n(\Vector{\theta})$ and $\bar{n}$ denote the (projected) number
density of galaxies in some direction $\Vector{\theta}$ and the mean
number density of galaxies, respectively. The quantity $\delta
n=n/\bar{n}-1$ is the projected number density contrast of the
galaxies.

Usually, the galaxies probed with $N(\Vector{\theta},\theta_{\rm ap})$
and those galaxies used to construct $M_{\rm
  ap}(\Vector{\theta},\theta_{\rm ap})$ maps are different; the latter
tend to be more distant as they probe the matter field containing the
galaxies used for $N$.  For that reason, we call the ``N-galaxies''
\emph{foreground galaxies} and the ``$M_{\rm ap}$-galaxies''
\emph{background galaxies}.

The projected galaxy number density contrast is related to the
three-dimensional galaxy number density contrast, $\delta_{\rm g}$,
via
\begin{equation}\label{glxydencontrast}
  \delta n(\Vector{\theta})=
  \int_0^{w_h}\d w\,p_{\rm f}(w)\,\delta_{\rm g}(f_{\rm
      K}(w)\Vector{\theta},w)
  \; .
\end{equation}
The function $p_{\rm f}(w)$ is the distribution of
foreground galaxies in comoving distance $w$ which will be estimated
from the observed distribution in (photometric) redshift. Note that
the galaxy distribution in redshift, $p^{\rm z}_{\rm f}(z)$, and distribution
in comoving distance are related by:
\begin{equation}
  p_{\rm f}(w)=p^{\rm z}_{\rm f}(z)\frac{\d z}{\d w}=p^{\rm z}_{\rm f}(z(w))\frac{H(z(w))}{c}\;,
\end{equation}
where $z(w)$ is the redshift as function of $w$.

Eq. \Ref{glxydencontrast} is the counterpart to Eq. \Ref{convergence}.
It is the projected density contrast of the galaxy density, while
$\kappa$ is the projected density contrast of the total matter.

\subsection{Aperture statistics and correlation functions}

In order to estimate the linear stochastic bias, Eq. \Ref{biascoeff},
using the aperture number count ($N$ related to $\delta_{\rm g}$) and
aperture mass statistics ($M_{\rm ap}$ related to $\delta_{\rm m}$) we
need to estimate the second-order moments of the aperture statistics,
i.e. \mbox{$\ave{N^n(\theta_{\rm ap})M^m_{\rm ap}(\theta_{\rm ap})}$}
with \mbox{$m+n=2$}.  There are two principal ways to estimate the
$2^{\rm nd}$-order moments of $N$ and $M_{\rm ap}$: either by placing
apertures at different positions onto the field (Hoekstra et al.
2001), or indirectly by estimating and transforming the two-point
correlation function of the galaxy number density, cosmic shear and
their cross-correlation (Hoekstra et al.  2002). In this paper, we are
going to use the latter method. How the correlation functions relate
to the aperture statistics will be summarised in the following.

\paragraph{Power spectra.} From the statistical point of view, the
joint $2^{\rm nd}$-order moments of the aperture statistics are
fluctuations, $\ave{N^2\left(\theta_{\rm ap}\right)}$ and
$\ave{M^2\left(\theta_{\rm ap}\right)}$, and correlations,
$\ave{N(\theta_{\rm ap})M_{\rm ap}(\theta_{\rm ap})}$, of smoothed
(statistically homogeneous and isotropic) random fields. They are
therefore auto- and cross-correlation power spectra seen through a
filter $[I\left(x\right)]^2$ (Hoekstra et al.  2002):
\begin{eqnarray}\label{aperture1}
  \ave{M_{\rm ap}^2\left(\theta_{\rm ap}\right)}&=&
  2\pi\int_0^\infty{\rm d}\ell\,\ell\,P_\kappa\left(\ell\right)
  \left[{I}\left(\ell\theta_{\rm ap}\right)\right]^2,
  \\
  \ave{N\left(\theta_{\rm ap}\right)M_{\rm ap}\left(\theta_{\rm ap}\right)}&=&
  2\pi\int_0^\infty{\rm d}\ell\,\ell\,P_{\rm n\kappa}\left(\ell\right)
  \left[{I}\left(\ell\theta_{\rm ap}\right)\right]^2,
  \\\label{aperture3}
  \ave{N^2\left(\theta_{\rm ap}\right)}&=&
  2\pi\int_0^\infty {\rm d}\ell\,\ell\,P_{\rm n}\left(\ell\right)
  \left[{I}\left(\ell\theta_{\rm ap}\right)\right]^2 ,
\end{eqnarray}
with the filter function
\begin{equation}
  {I}\left(x\right)\equiv
  \int_0^\infty {\rm d}s\,s\,u\left(s\right)
  J_0\left(s\,x\right)
  \; ,
\end{equation}
where 
\begin{eqnarray}
  (2\pi)^2\delta_{\rm D}(\Vector{\ell}+\Vector{\ell}^\prime)P_\kappa(|\Vector{\ell}|) &=&
 \langle\tilde{\kappa}(\Vector{\ell})\tilde{\kappa}(\Vector{\ell}^\prime)\rangle\;,\\
  (2\pi)^2\delta_{\rm D}(\Vector{\ell}+\Vector{\ell}^\prime)P_{\kappa\rm n}(|\Vector{\ell}|)&=&
  \langle\delta\tilde{n}(\Vector{\ell})\tilde{\kappa}(\Vector{\ell}^\prime)\rangle\;,\\
  (2\pi)^2\delta_{\rm D}(\Vector{\ell}+\Vector{\ell}^\prime)P_{\rm n}(|\Vector{\ell}|)&=&
  \langle\delta\tilde{n}(\Vector{\ell})\delta\tilde{n}(\Vector{\ell}^\prime)\rangle\;,
\end{eqnarray}
are the convergence auto-correlation, $P_\kappa$, convergence-galaxy
number density contrast cross-correlation, $P_{\kappa\rm n}$, and the
galaxy number contrast auto-correlation power spectrum, $P_{\rm n}$.
$\delta_{\rm D}$ denotes the Dirac delta function and a tilde is used
indicates the Fourier transform, such as
\begin{equation}
\tilde{\kappa}(\Vector{\ell})=\int\d^2\Vector{\theta}\,\kappa(\Vector{\theta})\,{\rm
  e}^{+{\rm i}\Vector{\theta}\Vector{\ell}}\;.
\end{equation}
We use $J_n(x)$ for the $n^{\rm th}$-order Bessel function of the
first kind.

Using Limber's equation in Fourier space (Kaiser 1992) we can derive
these power spectra from Eq. \Ref{glxydencontrast} and Eq.
\Ref{convergence}:
\begin{eqnarray}\label{autokappa}
  P_\kappa(\ell)&=&
  \int_0^{w_{\rm h}}\!\!\!\d w\frac{[\overline{W}(w)]^2}{[f_{\rm K}(w)]^2}\,
  P_{\rm m}\!\!\left(\frac{\ell}{f_{\rm K}(w)},w\right)\;,\\
  \label{kappan}
  P_{\kappa\rm n}(\ell)&=&
  \int_0^{w_{\rm h}}\!\!\!\d w\frac{\overline{W}(w)p_{\rm f}(w)}{[f_{\rm K}(w)]^2}\,
  (b\,r\,P_{\rm m})\!\!\left(\frac{\ell}{f_{\rm K}(w)},w\right)\;,\\
  \label{auton}
  P_{\rm n}(\ell)&=&
  \int_0^{w_{\rm h}}\!\!\!\d w\frac{[p_{\rm f}(w)]^2}{[f_{\rm K}(w)]^2}\,
  (b^2\,P_{\rm m})\!\!\left(\frac{\ell}{f_{\rm K}(w)},w\right)\;.
\end{eqnarray}
They will be needed below for the calculation of the calibration
factors.  Here, $P_{\rm m}(k,w)$ represents the 3D matter power
spectrum as a function of comoving distance $w$. In the above
equations, the (3D) galaxy-matter cross-power spectrum,
\begin{equation}
  (2\pi)^3\delta_{\rm D}(\Vector{k}+\Vector{k}^\prime)P_{\rm m,n}(k,w)=
  \ave{\tilde{\delta}_{\rm m}(\Vector{k},w)\tilde{\delta}_{\rm
      g}(\Vector{k}^\prime,w)}\; ,
\end{equation}
and (3D) galaxy power spectrum,
\begin{equation}
  (2\pi)^3\delta_{\rm D}(\Vector{k}+\Vector{k}^\prime)P_{\rm n}(k,w)=
  \ave{\tilde{\delta}_{\rm g}(\Vector{k},w)\tilde{\delta}_{\rm
      g}(\Vector{k}^\prime,w)}\; ,
\end{equation}
are expressed in terms of the linear stochastic bias in Fourier space:
\begin{equation}\label{truebias}
  b^2(k,w)\!=\!\frac{P_{\rm n}(k,w)}{P_{\rm m}(k,w)};
  r(k,w)\!=\!\frac{P_{\rm m,n}(k,w)}{\sqrt{P_{\rm n}(k,w)P_{\rm m}(k,w)}}\; .
\end{equation}
They are the Fourier space counterparts of the linear stochastic bias
parameters in Eq. \Ref{biascoeff}. \stressa{The Fourier space bias
  parameters are uniquely defined -- they are uniquely attached to
  Fourier modes which represent a certain physical scale --, whereas
  the real-space definition \Ref{biascoeff} requires further
  specification of a (smoothing) window function, $W_{\rm R}(x)$,
  through which the density fluctuations are observed.  Once the
  Fourier space bias is known and a window function is defined we can
  always work out the real-space bias parameters:
\begin{equation}
  b^2(R,\omega)=
  \frac{\int\d k\, k^2\,(b^2P_{\rm m})(k,\omega)|\widetilde{W}_{\rm R}(k)|^2}
       {\int\d k\,k^2\,P_{\rm m}(k,\omega)|\widetilde{W}_{\rm R}(k)|^2}
\end{equation}
and
\begin{equation}
  r(R,\omega)=\frac{1}{b(R,\omega)}
  \frac{\int\d k\, k^2\,(brP_{\rm m})(k,\omega)|\widetilde{W}_{\rm R}(k)|^2}
       {\int\d k\,k^2\,P_{\rm m}(k,\omega)|\widetilde{W}_{\rm R}(k)|^2}\;,
\end{equation}
where \mbox{$\widetilde{W}_{\rm R}(k)$} is the Fourier transform of
the window \mbox{$W_{\rm R}(x)$} with respect to $x$. Hence, all
information on the linear stochastic bias ($2^{\rm nd}$-order biasing) is
contained in the Fourier space bias parameters.}

Unbiased galaxies have $r=b=1$ for all $k$ and $w$. In general,
however, they are time- and scale-dependent. Note that for a very
narrow, or even delta function like $p_{\rm f}(w)$ Eq. \Ref{auton}
diverges because the assumptions made in the Limber equation break
down for those cases \stressa{(see Simon 2006)}.

\paragraph{Correlation functions.} The $2^{\rm nd}$-order statistics of
random fields is completely described by the power spectra.
Equivalently, we can consider correlations between pairs of points in
the random fields in real space which gives rise to the two-point
correlation functions. Although not carrying different information
than the power spectra, correlation functions are of great practical
value because they are relatively easily calculated; we just have to
consider pairs of galaxies in our case.

These correlators are here: a) the angular correlation of the
foreground galaxy positions, $\omega\left(\theta\right)$, b) the mean
tangential shear about foreground galaxies, $\ave{\gamma_{\rm
    t}}\left(\theta\right)$, and c) the shear-shear correlations
$\xi_\pm\left(\theta\right)$ as determined from the ellipticities of
the background galaxies (Hoekstra et al.  2002):
\begin{eqnarray}\label{twopointcorr1}
  \omega\left(\theta\right)&=&
  \ave{\delta n\left(\theta+x\right)\delta n\left(x\right)}
  \\\nonumber &=&
  \int_0^\infty \frac{{\rm d}s~s}{2\pi} P_{\rm n}\left(s\right)J_0\left(s\theta\right),
  \\\label{twopointcorr2}
  \ave{\gamma_{\rm t}}\left(\theta\right)&=&
  \ave{\delta n\left(\theta+x\right)\gamma_{\rm t}\left(x\right)}
  \\ \nonumber
  &=&
  \int_0^\infty \frac{{\rm d}s~s}{2\pi} P_{\kappa\rm n}\left(s\right)J_2\left(s\theta\right),
  \\\label{twopointcorr3}
  \xi_\pm\left(\theta\right)&=&
  \ave{\gamma_{\rm t}\left(\theta+x\right)\gamma_{\rm
      t}\left(x\right)}
  \pm\ave{\gamma_\times\left(\theta+x\right)\gamma_\times\left(x\right)}\\\nonumber
  &=&
  \int_0^\infty \frac{{\rm d}s~s}{2\pi} P_\kappa\left(s\right)J_{0,4}\left(s\theta\right)
  \; .
\end{eqnarray}
The correlator $\omega(\theta)$ is a measure for the probability of
finding a galaxy at a separation $\theta$ from another galaxy. The
galaxy-galaxy lensing, $\ave{\gamma_{\rm t}}(\theta)$, is the mean
tangential shear around a foreground galaxy at a separation $\theta$,
and the two-point shear-shear correlations, $\xi_\pm(\theta)$, quantify
the correlations of the cross- and tangential shear components
relative to the line connecting two background galaxies with an
angular separation $\theta$. All correlators are linearly related to
the corresponding power spectra.

\paragraph{Transformation integrals.} The relations between the
correlators and the power spectra, Eqs.
\Ref{twopointcorr1}-\Ref{twopointcorr3}, can be inverted with respect
to the power spectra (Schneider et al. 2002; Hoekstra et al. 2002), so
that invoking Eqs. \Ref{aperture1}-\Ref{aperture3} we can express the
aperture moments in terms of the two-point correlators:
\begin{eqnarray}\label{integraltrans1} 
  \ave{M_{\rm ap}^2\left(\theta_{\rm ap}\right)}&=&
  \frac{1}{2}
  \int_0^\infty\!\!{\rm d}x~x~\left[\xi_+\left(x\theta_{\rm
        ap}\right)T_+\left(x\right)\right.\\\nonumber
  &+&\left.\xi_-\left(x\theta_{\rm
        ap}\right)T_-\left(x\right)\right],\\
\label{integraltrans1b} 
  \ave{N\left(\theta_{\rm ap}\right)M_{\rm ap}\left(\theta_{\rm ap}\right)}&=&
  \int_0^\infty\!\!{\rm d}x~x~\ave{\gamma_{\rm   t}}\left(x\theta_{\rm ap}\right)
  F\left(x\right),   \\\label{integraltrans2} 
  \ave{N^2\left(\theta_{\rm ap}\right)}&=&\int_0^\infty\!\!
  {\rm d}x~x~\omega\left(x\theta_{\rm ap}\right) T_+\left(x\right),          
\end{eqnarray}
where we introduced the auxiliary functions
\begin{eqnarray}\label{tpmtrans}
  T_{+,-}\left(x\right)&\equiv&            \left(2\pi\right)^2\int_0^\infty
  {\rm d}s~s~\left[
    I\left(s\right)\right]^2J_{0,4}\left(sx\right), 
  \\\label{ftrans}
  F\left(x\right)&\equiv&        \left(2\pi\right)^2       \int_0^\infty
  {\rm d}s~s~\left[  I\left(s\right)\right]^2J_2\left(sx\right)\;.
\end{eqnarray}
The Eqs. \Ref{integraltrans1}-\Ref{integraltrans2} are the basis for
the method that is used in this paper. The two-point correlators are
estimated from the data and then afterwards integrated in order to
obtain the $2^{\rm nd}$-order moments of the joint PDF of aperture
mass and aperture number count.

If we substitute in the foregoing Eq. \Ref{integraltrans1} and
\Ref{integraltrans1b} tangential shear components, $\gamma_{\rm t}$,
by cross shear components, $\gamma_\times$, and vice versa then we
obtain the transformation integrals for the corresponding B-modes; we
then have $\ave{\gamma_\times}$ in \Ref{integraltrans1b} and the $\xi_-$
 term in \Ref{integraltrans1} changing sign.

\section{Data}
\label{gabodssection}

Here we will give only a brief account of the GaBoDS. For details
concerning the GaBoDS, its data reduction and catalogue creation, we
refer the reader to Schirmer (2004), Erben et al. (2005) and in
particular Hetterscheidt et al. (2006).

\subsection{The GaBoDS fields and their reduction}

 \begin{figure}
   \begin{center}
     \psfig{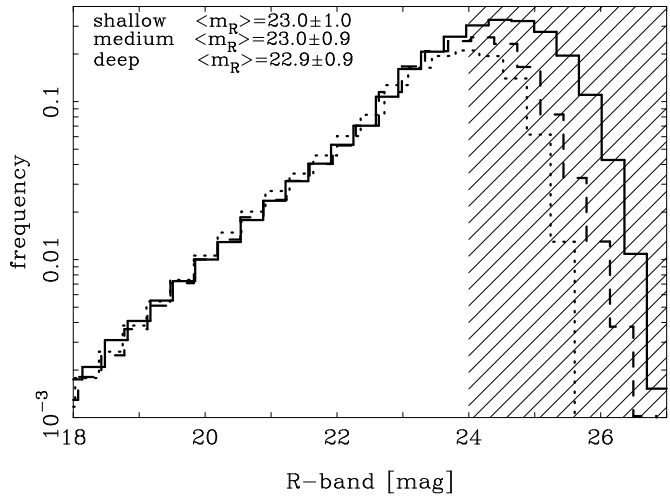}
   \end{center}
   \caption{\label{maghist} Frequency of apparent galaxy $R$-band
     magnitudes in the shallow, medium and deep part of GaBoDS
     (foreground galaxy samples). The distribution functions have been
     normalised by the area between $18\le R\le24\,\rm mag$ since
     galaxies fainter than $24\,\rm mag$ are not considered in this
     paper. As can be seen in this plot, all three data sets have
     roughly comparable distributions for $R\le 24\,\rm mag$.}
 \end{figure}

 The GaBoDS comprises roughly $18.6~\rm deg^2$ of high-quality data
 (seeing better than one arcsec) in $R$-band taken with the Wide Field
 Imager (WFI) mounted on the 2.2m telescope of MPG/ESO at La Silla,
 Chile; the \mbox{$33^\prime\times 34^\prime$} field-of-view is
 covered with 8 CCD chips.  Due to the dither pattern applied, the
 effective field-of-view can be as large as roughly
 $40^\prime\times40^\prime$.  The data set was compiled mostly from
 archival ESO data, for which the archive utility \texttt{querator}
 (Pierfederici 2001) has been developed, together with about four
 square degree coming from our own observations.  The positions of the
 fields were chosen randomly from regions of small stellar densities
 at high galactic latitudes.  The limiting magnitudes of the fields is
 inhomogeneous, ranging between $25.0\,\rm mag$ and $26.5\,\rm mag$
 ($5\sigma$ in a $2^\pprime$ aperture radius) in the $R$-band
 depending on the exposure time and on the fraction of time the seeing
 was acceptable for gravitational lensing applications. The data set
 can roughly be categorised into a shallow ($t\le 7\,\rm ksec$, total
 $9.6\,\rm~deg^2$), medium ($7\,{\rm ksec}<t\le10\,{\rm ksec}$, total
 $7.4\,\rm deg^2$) and deep ($10\,{\rm ksec}<t\le 56\,\rm~ksec$, total
 $2.6\,\rm~deg^2$) set depending on the total usable integration time
 $t$ for each field.

 The data imposed new, high demands on the data reduction, which
 resulted into the development of a data reduction pipeline whose
 usage is not restricted to the aforementioned instrument only; it has
 successfully been tested on data from various other instruments
 (Erben et al. 2005).

 For the final analysis, we consider the shallow, medium and deep part
 of the GaBoDS comprising in total $52$ WFI fields corresponding to an
 area of about $15~\rm deg^2$. We rejected nine fields: CAPO and all
 fields belonging to the C0 series (eight), as the quality of the PSF
 correction was decided to be not sufficient enough for weak
 gravitational lensing applications (see Hetterscheidt et al.  2006).
 As will be explained shortly we do not consider galaxies, for both
 lensing and foreground object catalogues, that are fainter in the
 $R$-band than $24\,\rm mag$.  Applying this cut at the faint end
 makes the GaBoDS categories shallow, medium and deep roughly
 comparable with each other as can be seen by the magnitude histogram
 in Fig.  \ref{maghist}, and it allows us to estimate the redshift
 distribution of galaxies (see below).

\subsection{Selection of lensing catalogues}

 \begin{figure}
  \begin{center}
     \epsfig{file=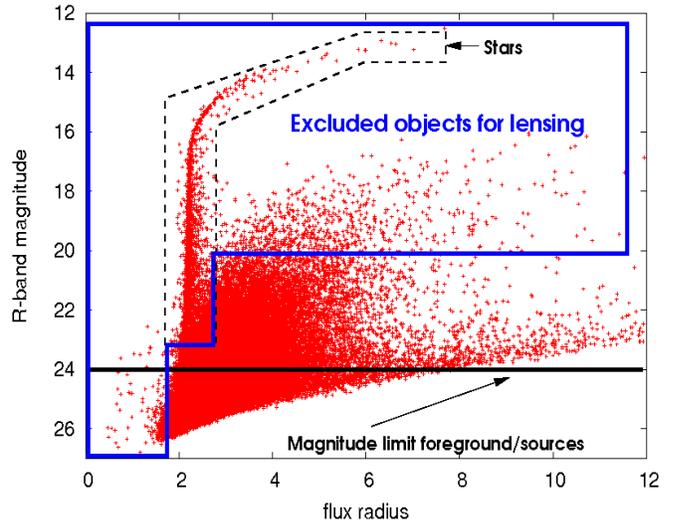,width=90mm}
   \end{center}
   \caption{\label{cut}Magnitude vs. half-light radius plot of objects
     found by \texttt{SExtractor} in one particular field.  Stars
     appear as almost vertical branch and can be separated from
     galaxies with high confidence. The solid and dashed box roughly
     encircles objects excluded for the lensing catalogue (Schirmer et
     al. 2003).}
 \end{figure}

 After the data reduction process, \texttt{SExtractor}
 (Bertin~\&~Arnouts 1996) was used to compile a catalogue of
 \emph{source galaxy candidates} needed for the cosmic shear analysis.
 For the rather conservative selection of source candidates, the final
 co-added science frames are first smoothed with a Gaussian kernel of
 $2.5$ pixel FHWM. One pixel corresponds to $0.\!\!^\pprime238$. A source
 candidate further needs to consist of at least $5$ contiguous pixels
 with a total flux greater than $1.5\,\sigma$ above the background
 noise level, and it has to possess a clearly defined quadrupole
 moment ($cl\ne 0$, \texttt{analyseldac}) and centroid.  Stars and
 galaxies are distinguished in a magnitude vs.  half-light radius plot
 of the selected objects (see Fig.  \ref{cut} for an example).  In
 this scatter plot, stars that are not too faint are clearly
 identified as a column of objects with roughly identical half-light
 radius $r_\star$.  Objects with a half-light radius smaller than
 $r_\star$ are rejected as source candidates.  An exception are
 objects in the faint part (fainter than $23.5~\rm mag$ in $R$-band)
 near this column.

 As accurate measurements of galaxy shapes are the key in a weak
 lensing analysis, the quadrupole moments in the galaxy light profiles
 of the source candidates have to be corrected for PSF effects:
 atmospheric turbulence and instrumental effects also distort the
 galaxy images. This is done using the KSB method (Kaiser et al.
 1995).  A detailed description of the PSF correction procedure may be
 found in Erben et al. (2001), Heymans et al. (2006) or Hetterscheidt
 et al. (2006).  The PSF fitting polynomial used is of order two or
 three.

 In the estimators of the aperture statistics, every source galaxy is
 weighted with a statistical weight.  This weight, $w_i$, is defined
 by the variance $\sigma^2_\epsilon$ in ellipticity of the $12$
 nearest neighbours of a galaxy $i$ in the magnitude vs. half-light
 radius diagram: \mbox{$w_i=1/(\sigma^2_\epsilon+\bar{\sigma}^2)$},
 where \mbox{$\bar{\sigma}^2=0.16$} is the variance of the unlensed
 galaxies. In the case that the PSF corrected ellipticity of a galaxy
 exceeds $\left|\epsilon\right|=1.0$ it automatically is attributed
 the weight zero and is hence not considered further in the analysis.
 Applying this cut removes rare outliers with unrealistic
 ellipticities, produced by the KSB technique. The final lensing
 catalogue is split into three magnitude bins BACK, BACK-II and
 BACK-III, see Table \ref{gabodstable}.

\subsection{Selection of foreground objects}

The actual foreground objects, of which the bias parameters are
measured, are selected with the same \texttt{SExtractor} parameters as
the galaxies candidates in the lensing catalogue. Galaxies are finally
selected from this catalogue via a manually defined box in the
magnitude/half-light radius diagram around the star branch.

\begin{table}
  \begin{center}
    \caption{\label{gabodstable}The table lists the limits of the
      magnitude bins, the total number of objects for all $52$ fields
      (deep, medium and shallow fields in GaBoDS), the mean
      redshift and the $1\sigma$-variance inside each bin.}
      \begin{tabular}{llcc} 
        \multicolumn{4}{c}{\textbf{Foreground object  catalogue}}\\\\ 
        sample&bin limits [mag]&\#objects&$\langle z\rangle$\\ \hline\\
        FORE-I & $19.5\le R<21.0$ & $6.5\times10^4$ & $0.34\pm0.18$ \\
        FORE-II & $21.0\le R<22.0$ & $1.2\times10^5$ &$0.47\pm0.22$\\ 
        FORE-III & $22.0\le R<23.0$ &$2.5\times10^5$ &  $0.62\pm0.27$\\\\\\
        \multicolumn{4}{c}{\textbf{Background source catalogue}}\\\\
        sample&bin limits [mag]&\#objects&$\langle z\rangle$\\ \hline\\
        \textbf{BACK} & $21.5\le R<24.0$ & $6.2\times10^5$ &
        $0.67\pm0.29$ \\\\ 
        BACK-II & $22.0\le R<24.0$ & $5.5\times10^5$ &
        $0.69\pm0.28$ \\ 
        BACK-III & $23.0\le R<24.0$ & $3.5\times10^5$ & 
        $0.74\pm0.28$
      \end{tabular}
  \end{center}
\end{table}

In order to select for the bias analysis different mean redshifts of
the (foreground) object catalogues, we subdivide the object catalogue
into the three different $R$-band bins \mbox{FORE-I}, \mbox{FORE-II}
and \mbox{FORE-III} as stated in Table \ref{gabodstable}.

 \begin{figure*}
   \begin{center}
     \psfig{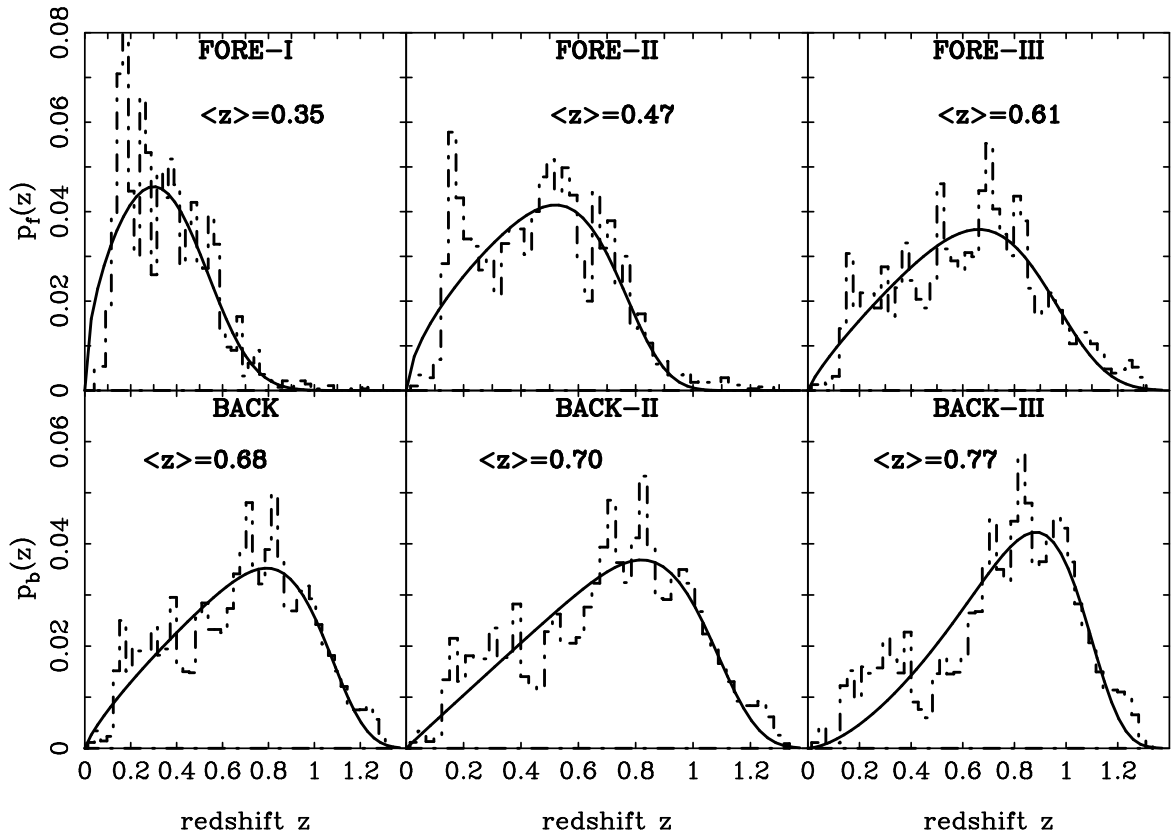}
   \end{center}
   \caption{\label{gabodspofz} Redshift distribution of the foreground
     and background galaxies as estimated from the photometric
     redshifts in the \mbox{COMBO-17} fields A901, AXAF (CDFS) and S11
     (dashed doted lines); the histograms are not normalised to unity.
     The solid lines are maximum-likelihood fits of Eq.
     \Ref{brainerd} to the histograms.}
 \end{figure*}

\subsection{Distribution in redshift of the galaxy samples}
\label{pzsect}

To estimate the redshift distribution of the galaxies -- both
foreground objects and background sources -- we average the
photometric redshift distribution in the different magnitude bins of
the fields A901, AXAF and S11 (see Fig.  \ref{gabodspofz}).  These
three fields are contained in the deep part of the GaBoDS and were
observed as part of the \mbox{COMBO-17} survey (Wolf et al. 2004) in
$17$ colours yielding quite accurate photometric redshifts with an
uncertainty of \mbox{$\delta z\approx 0.02~(1+z)$} for objects
brighter than \mbox{$R=23~\rm mag$}. Less accurate but still available
are photometric redshifts for objects with \mbox{$23<R<24$}. The
photometric redshift distribution of the \mbox{COMBO-17} galaxies is
assumed to be representative for our whole catalogue.

High-redshift galaxies with \mbox{$z>1.4$} are ``missing'' in the
COMBO-17 sample because they were reassigned a redshift
$z<1.4$. Recently, in Coe et al. (2006), the Hubble Ultra-Deep Field
has been used to validate the photometric redshifts in COMBO-17. It
has been found that the agreement is good for $R\lesssim23.7$ and
especially tight for $R<23$.  We conclude, therefore, that we have got
a reliable estimate of the redshift distribution in our samples.

For the source galaxies carrying the $M_{\rm ap}$-signal, the
magnitude bin BACK is used throughout. As can be seen in Table
\ref{gabodstable}, by varying only the lower limit, but keeping the
upper limit of the magnitude bin fixed to \mbox{$R=24~\rm mag$}, one
cannot shift the mean of the background redshift distribution to much
higher values than \mbox{$z\approx0.7$}; essentially, only the number
of sources in the bin decreases. A large mean redshift of the source
galaxies is desired to achieve a good lensing efficiency but more
important is a large number of galaxies to achieve a good
signal-to-noise ratio. Since we do not use objects fainter than
\mbox{$24~\rm mag$} in order to maintain good accuracy in the estimate
for the redshift distribution of the background and to have a roughly
homogeneous data set, the bin BACK for all three foreground bins
\mbox{FORE-I}, \mbox{FORE-II} and \mbox{FORE-III} is the best choice.

The COMBO-17 sample used to estimate the redshift distribution in the
galaxy sub-samples is relatively small. Clearly, it has features --
large galaxy clusters or voids -- which are not representative for the
whole GaBoDS sample. For example, consider the peaks at low redshift
in the foreground samples, Fig. \ref{gabodspofz}. In order to have a
smoother, more representative distribution we fit an empirical
redshift distribution to the COMBO-17 histograms:
\begin{equation}\label{brainerd}
  p(z)=
  \frac{1}{\Gamma(\frac{1+\alpha}{\beta})z_0}
  \left(\frac{z}{z_0}\right)^\alpha\,
  \exp{\left(-\left[\frac{z}{z_0}\right]^\beta\right)}\;.
\end{equation}
The best-fits are shown in Fig. \ref{gabodspofz}. They are used in the
analysis further on instead of the COMBO-17 histograms. The best fit
parameters are compiled in Table \ref{pofzfits}. \stressa{These
  parameters are highly degenerate for which reason their statistical
  errors are not given in the table. However, statistical
  uncertainties of the mean redshift in the different samples and
  uncertainties of the galaxy bias parameters originating from
  uncertainties in $p(z)$ will be given in the following.}

\stressa{Clearly, the estimated redshift distributions still suffer
  from cosmic variance errors because the COMBO-17 survey area is with
  \mbox{$\sim0.75\,\rm deg^2$} relatively small. In order to get an
  estimate for the statistical uncertainties due to cosmic variance in
  the samples' redshift distribution we use the widely applied
  Jackknife method: The photometric redshift distributions of merely
  two of the three fields are combined. With three ways of combining
  this yields overall \mbox{$N=3$} Jackknife samples.  To estimate the
  standard deviation of the mean redshift, $\bar{z}$, one computes
  from each Jackknife sample the mean redshift, $\bar{z}_i$.
  According to the Jackknife method the statistical $1\sigma$-error of
  the mean is then roughly:
  \begin{equation}\label{jackknife}
    \sigma^2(\bar{z})=\frac{N-1}{N}\sum_i(z_i-\bar{z})^2\;,
  \end{equation}
  where $\bar{z}$ is the mean redshift obtained by combining all three
  COMBO-17 redshift distributions. The results for $\sigma(\bar{z})$
  are listed in Table \ref{pofzfits}. As can be seen there the
  uncertainty of $\bar{z}$ ranges from
  \mbox{$\sigma(\bar{z})/\bar{z}\approx10\%$} to
  \mbox{$\sigma(\bar{z})/\bar{z}\approx2\%$} for \mbox{FORE-I} to
  \mbox{BACK}, respectively. This behaviour makes sense because the
  number of galaxies increases when going from the shallower to the
  deeper samples.}

\stressa{The problem of the calibration of redshift distributions for
  cosmic shear studies has recently been studied by van Waerbeke et
  al. (2006). They find a statistical uncertainty of
  \mbox{$\sigma(\bar{z})=0.03-0.04$} for a $0.75\,\rm deg^2$ survey
  with mean \mbox{$\bar{z}\sim1$}. This value is somewhat higher than
  our estimate.}

\stressa{The Jackknife samples can also be used to assess how the
  statistical uncertainty of the full $p(z)$'s translates into the
  inferred galaxy bias parameters. This problem will be addressed in
  Sect.  \ref{calibrationsection}.}

\begin{table}
  \caption{\label{pofzfits}Best-fit parameters of the template
    redshift distribution, Eq. \Ref{brainerd}, to the COMBO-17
    histograms. $\bar{z}$ is the mean of the template redshift
    distribution. \stressa{The statistical errors of $\bar{z}$ are derived from
    the field-to-field variance in COMBO-17.}}
\begin{center}
\begin{tabular}{lcccc}
  galaxy sample&$z_0$&$\alpha$&$\beta$&$\bar{z}$\\\hline\\
  FORE-I&0.534&0.509&3.173&$0.35\pm0.03$\\
  FORE-II&0.765&0.617&5.839&$0.47\pm0.03$\\
  FORE-III&0.945&0.830&5.103&$0.61\pm0.02$\\\\
  \textbf{BACK}&1.069&0.809&7.369&$0.68\pm0.02$\\
  BACK-II&1.072&0.988&7.655&$0.70\pm0.02$\\
  BACK-III&1.073&1.611&8.560&$0.77\pm0.02$
\end{tabular}
\end{center}
\end{table}

\section{Outline of the method}
\label{methodsection}

The approach to obtain the bias parameters from lensing adopted here
proceeds in several steps: 
\begin{enumerate}
\item estimating the binned correlators $\omega(\theta)$,
  $\ave{\gamma_{\rm t}}(\theta)$ and $\xi_\pm(\theta)$ in all
  individual survey fields,
\item numerical integration of the correlators to obtain
  \mbox{$\ave{N^m(\theta_{\rm ap})M_{\rm ap}^n(\theta_{\rm ap})}$} for
  \mbox{$m+n=2$} (E-modes and B-modes),
\item repetition of 1. and 2. with bootstrapped data sets to obtain
  statistical errors of the aperture statistics in the single fields,
\item combining the individual field measurements and evaluating the
  bias parameters as a function of aperture radius from the combined
  signal (includes calibration),
\item bootstrapping of the combined signal to estimate the error in
  the final signal and the covariances between the different bins.
\end{enumerate}
A detailed account of these steps is given in the following.

\subsection{Practical estimators for the correlators}
\label{usingtwoptcorrelators}

The correlators are estimated by using 
\begin{eqnarray}\label{omega}
  \omega\left(\theta\right)&=&\frac{DD}{RR}-2\frac{DR}{RR}+1,
  \\\label{glxglxlensing}
  \ave{\gamma_{\rm   t}}\left(\theta\right)&=&  \frac{\sum_{i,j}^{N_{\rm
        f},N_{\rm                                               b}}\epsilon_{{\rm
        t},i}w_i\Delta_{ij}\left(\theta\right)}{\sum_{i,j}^{N_{\rm    f},N_{\rm
        b}}w_i\Delta_{ij}\left(\theta\right)},
  \\
  \xi_\pm\left(\theta\right)&=&\frac{\sum_{i,j}^{N_{\rm
        b}}w_iw_j\Delta_{ij}\left(\theta\right)              \left(\epsilon_{{\rm
          t},i}\epsilon_{{\rm
          t},j}\pm\epsilon_{{\times},i}\epsilon_{{\times},j}\right)}
  {\sum_{i,j}^{N_{\rm
        b}}w_iw_j\Delta_{ij}\left(\theta\right)}\;,
\end{eqnarray}
where
\begin{equation}
  \Delta_{ij}\left(\theta\right)\equiv\left\{
    \begin{array}{ll}
1&{\rm for}~\theta\le\left|\theta_i-\theta_j\right|<\theta+\delta\theta\\      0&
{\rm otherwise}
    \end{array}\right.  \; .
\end{equation}
The $w_i$ are statistical weights of the individual galaxies which are
used to account for the fact that the values of the image
ellipticities, $\epsilon_i$, of the galaxies do not have all the same
accuracy. Ellipticities of fainter and smaller galaxies are determined
with a lower accuracy than for larger and brighter galaxies. $N_{\rm
  f}$ and $N_{\rm b}$ are the number of foreground and background
galaxies; $\epsilon_{{\rm t},i/j}$ and $\epsilon_{\times, i/j}$ are
the tangential and cross ellipticity components relative to the line
connecting the galaxy pair $i,j$.

The estimator of the spatial correlation $\omega(\theta)$ has been
introduced by Landy~\&~Szalay (1993). It requires to count the number
of galaxy pairs with a separation between $\theta$ and
\mbox{$\theta+\delta\theta$}, namely the number of pairs in the data
(foreground galaxies), denoted by $DD$, the number of pairs in a
random mock catalogue, $RR$, and the number of pairs that can be
formed with one data galaxy and one mock data galaxy, $DR$. The random
mock catalogue is computed by randomly placing the galaxies, taking
into account the geometry of the data field, i.e. by avoiding
outmasked regions. We make $40$ random galaxy catalogues and average
the pair counts obtained for $DR$ and $RR$.

In an estimate of $\omega(\theta)$, Eq. \Ref{omega}, there is always
an uncertainty about the mean galaxy density $\bar{n}$ which is the
larger the smaller the area of the field under consideration. This
introduces a bias known as the integral constraint (Groth~\&~Peebles
1977), that systematically reduces the angular correlation,
\mbox{$\omega(\theta)\mapsto\omega(\theta)-C$}, by a constant value
\mbox{$C>0$}.  As pointed out by Hoekstra et al. (2002) $\ave{N^2}$ is
independent of the integral constraint when the aperture filter $u$
is, as in our analysis, compensated because
\begin{equation}
  \int_0^\infty\d x\,x\,C\,T_+(x)=C\,\int_0^\infty\d x\,x\,T_+(x)=0
  \; . 
\end{equation}
Therefore the estimator bias $C$ does not make any contribution to
$\ave{N^2}$ and does not need to be corrected. This nice feature makes
the aperture statistics with a compensated filter a convenient tool to
study galaxy clustering even outside a weak lensing context. We will
briefly come back to this point in Sect. \ref{resultsection}.

Concerning the estimator for mean tangential shear, $\ave{\gamma_{\rm
    t}}(\theta)$, all pairs of foreground and background galaxies
within separation between $\theta$ and $\theta+\delta\theta$ have to
be considered; $\epsilon_{{\rm t},i}$ is the tangential ellipticity
component of the background galaxy with respect to the line connecting
the foreground and background galaxy.  Similarly, for
$\xi_\pm(\theta)$ all pairs of background galaxies within some
separation interval are considered.

For the GaBoDS analysis, we bin the three correlators into $800$
logarithmic bins spanning a range between
\mbox{$0.\!\!^\pprime05<\theta\le 48^\prime$} (the diagonal of a
single WFI field).  In order to reduce the computation time for the
correlations, a binary tree data structure as in Pen~\&~Zhang (2003),
Moore et al. (2001) or Jarvis et al. (2004) is used.

\subsection{Aperture filter}

To weight density fluctuations inside apertures we use a compensated
polynomial filter (Hoekstra et al. 2002, 2001; Schneider et al.  1998)
\begin{equation}
  \label{filter}
  u\left(x\right)=
  \frac{9}{\pi}\left(1-x^2\right)\left(\frac{1}{3}-x^2\right)\,H(1-x)\;,
\end{equation} 
which by definition vanishes for $x\ge 1$; $H(x)$ denotes the
Heaviside step function.  The filter has the effect that only matter
or galaxy number density fluctuations from a small range of angular
scales contribute to the $N$- or $M_{\rm ap}$-signal; it acts as a
narrow-band filter for the angular modes with highest sensitivity to
$\ell_{\rm c}\sim4.25/\theta_{\rm ap}\approx
0.68\times2\pi/\theta_{\rm ap}$.  Apertures with radius
$\theta_{\rm ap}$ therefore effectively probe a comoving physical
scale of $f_{\rm K}(\bar{w})\,\theta_{\rm ap}/0.68$, if $\bar{w}$ is the
median comoving radial distance of the galaxy sample under
examination.

All auxiliary functions \Ref{tpmtrans}-\Ref{ftrans}, which are
required for transforming the correlators, vanish outside the interval
$x\in\left[0,2\right]$ due to the finite support of $u$. This reduces
the transformation integrals \Ref{integraltrans1}-\Ref{integraltrans2}
to a finite integration range $[0,2\theta_{\rm ap}]$.  Therefore, with
square $30^\prime\times 30^\prime$ WFI fields we are able to estimate
the aperture moments out to about
$\frac{1}{2}\sqrt{2}\,30^\prime\approx21^\prime$.

\subsection{Calibration of bias parameters}
\label{calibrationsection}

 \begin{figure*}
   \begin{center}
     \psfig{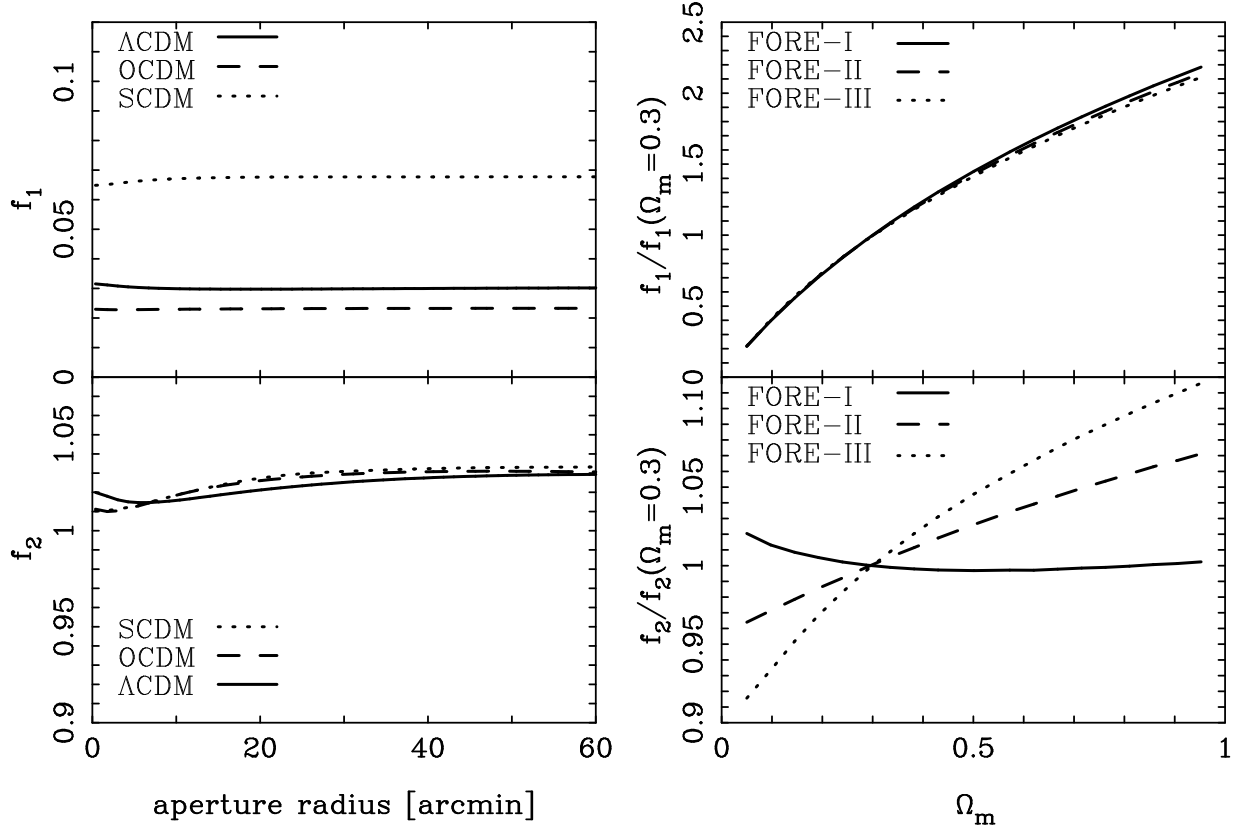}
   \end{center}
   \caption{\label{calibrationfig}\textbf{Left}: The two figures show
     the scale-dependence of the calibration factors $f_{1/2}$, for
     sample FORE-I only, for three different fiducial cosmologies;
     SCDM (dotted): $\Omega_{\rm m}=1.0$, $\Omega_\Lambda=0$;
     $\Lambda\rm CDM$ (solid): $\Omega_{\rm m}=0.3$,
     $\Omega_\Lambda=0.7$; OCDM (dashed): $\Omega_{\rm m}=0.3$,
     $\Omega_\Lambda=0$.  \textbf{Right}: These plots were obtained by
     averaging $f_{1/2}$ over a range of aperture radii,
     \mbox{$1^\prime\le\theta_{\rm ap}<60^\prime$}, assuming different
     fiducial cosmologies. For all cosmologies, $\Omega_{\rm m}$ is the
     only free parameter. The others are:
     $\Omega_\Lambda=1-\Omega_{\rm m}$, $\Gamma=\Omega_{\rm m}h$,
     $\sigma_8=0.41\,\Omega_{\rm m}^{-0.56}$ and $h=0.7$. The average
     values for $f_{1/2}$ in this figure are divided by
     $f_{1/2}(\Omega_{\rm m}=0.3)$, the here adopted calibration. For
     $p_{\rm f}(z)$, we have \mbox{FORE-I} (solid), \mbox{FORE-II}
     (dashed) and \mbox{FORE-III} (dotted); $p_{\rm b}(z)$ is as in
     BACK.}
 \end{figure*}

 To summarise, the aperture mass $M_{\rm ap}$, Eq.
 \Ref{aperturemassdef}, is proportional to the (weighted) projected
 total matter density contrast $\delta_{\rm m}$, whereas the aperture
 number count $N$, Eq.  \Ref{aperturecountdef}, is proportional to the
 number density contrast of the galaxy distribution $\delta_{\rm g}$.
 Both aperture measures are defined on some scale by the filter
 function $u$ and the aperture size $\theta_{\rm ap}$.  This is
 exactly what we need to study the biasing of the galaxy distribution
 with respect to the matter distribution, as has been pointed out by
 Schneider (1998) and van Waerbeke (1998).  Therefore, we can define
 biasing parameters in analogy to Eqs.  \Ref{biascoeff} (Hoekstra et
 al. 2002)
\begin{eqnarray}
\label{weaklinbias}  
b\left(\theta_{\rm  ap}\right)&=&
f_1\left(\theta_{\rm  ap}\right) \sqrt{\frac{\ave{N^2\left(\theta_{\rm
ap}\right)}}{\ave{M^2_{\rm ap}\left(\theta_{\rm
ap}\right)}}}\;,\\\nonumber 
r\left(\theta_{\rm ap}\right)&=&
f_2\left(\theta_{\rm ap}\right)\frac{\ave{N\left(\theta_{\rm
ap}\right)M_{\rm ap}\left(\theta_{\rm ap}\right)}}
{\sqrt{\ave{N^2\left(\theta_{\rm ap}\right)}\ave{M^2_{\rm
ap}\left(\theta_{\rm ap}\right)}}}
 \; .
\end{eqnarray}
The three-dimensional number density that $N$ is sensitive to covers
in general not the same volume that is probed by $M_{\rm ap}$ ($p_{\rm
f}(w)$ versus $\overline{W}(w)$). Naively identifying $N$ with the
galaxy number density contrast, $\delta_{\rm g}$, and $M_{\rm ap}$
with the matter density contrast, $\delta_{\rm m}$ therefore gives the
wrong bias parameters.  \stressa{This $N/M_{\rm ap}$-mismatch in
sensitivity to density fluctuations at the same radial distance thus
requires to make a correction of the bias parameters which is done by
means of the calibration factors $f_{1/2}$.}

According to Hoekstra et al. (2002), the calibration factors have to
be calculated based on some theoretical $P_{\rm m}(k,w)$ by means of
 \begin{eqnarray}
   \label{calibration} 
   f_1\left(\theta_{\rm ap}\right)&=&
   \left.\sqrt{\frac{\ave{M^2_{\rm       ap}\left(\theta_{\rm      ap}\right)}}
       {\ave{N^2\left(\theta_{\rm                     ap}\right)}}}\right|_{r=b=1}\; ,\\\nonumber
   f_2\left(\theta_{\rm                                      ap}\right)&=&
   \left.\frac{\sqrt{\ave{N^2\left(\theta_{\rm          ap}\right)}\ave{M^2_{\rm
             ap}\left(\theta_{\rm       ap}\right)}}}      {\ave{N\left(\theta_{\rm
             ap}\right)M_{\rm ap}\left(\theta_{\rm ap}\right)}}\right|_{r=b=1} \; ,
 \end{eqnarray} 
 where \mbox{$\ave{N^n(\theta_{\rm ap})M^m_{\rm ap}(\theta_{\rm
       ap})}$}, \mbox{$n+m=2$}, in these equations have to be
 evaluated by Eqs.  \Ref{aperture1}-\Ref{aperture3} and
 \Ref{autokappa}-\Ref{auton}, specifically for the redshift
 distributions of foreground, $p_{\rm f}(z)$, and background galaxies,
 $p_{\rm b}(z)$, in the data and for a fiducial cosmological model
 \emph{assuming that galaxies are not biased with respect to the dark
   matter}, \emph{i.e.}  $b(k,w)=r(k,w)=1$.

 Importantly, it turns out (van Waerbeke 1998) that the calibration
 factors $f_1$ and $f_2$ vary only slightly, mostly on scales below
 $\theta_{\rm ap}\lesssim5^\prime$, for realistic aperture radii
 $\theta_{\rm ap}$ within a fixed fiducial cosmological model. This is
 strictly true if the dark matter power spectrum can be described by a
 power law, or -- since we are, for a fixed aperture radius, sensitive
 to only a very localised range in $\ell$-space due to the adopted
 aperture filter -- if the power spectrum is approximately a power law
 over the selected range in Fourier space.

 For examples, see Fig.  \ref{calibrationfig} (upper left and bottom
 left) where $f_{1/2}$ are plotted for three fiducial cosmological
 models assuming the redshift distribution of \mbox{FORE-I} and BACK.
 The calibration factors show very little dependence on $\theta_{\rm
   ap}$.  Hence, a scale-dependence of the uncalibrated measurements
 immediately indicates a real scale-dependence in the bias parameter
 \emph{without} fixing the fiducial cosmology!  Moreover, it means
 that the calibration factors can be worked out for the linear or
 quasi-linear regime which is understood much better than the
 non-linear regime. Still, when calibrating our measurements we also
 take into account the dependence on scale.

 We calculated the calibration factors $f_{1/2}$ for a range of
 spatially flat fiducial cosmologies, \mbox{$\Omega_{\rm
     m}+\Omega_\Lambda=1$}, using the redshift distribution in our
 data set (right column in Fig.  \ref{calibrationfig}), assuming
 constraints on \mbox{$\sigma_8\propto\Omega_{\rm m}^{-0.56}$} from
 cluster abundances (White et al. 1993) and the shape parameter
 \mbox{$\Gamma=\Omega_{\rm m}h$} for a negligible baryon density
 \mbox{$\Omega_{\rm b}\approx0$} (Efstathiou et al.  1992).  The
 relation between $\sigma_8$ and $\Omega_{\rm m}$ is scaled such that
 \mbox{$\sigma_8=0.8$} corresponds to \mbox{$\Omega_{\rm m}=0.3$}.
 This value of $\sigma_8$ for the power spectrum normalisation is
 suggested by the GaBoDS data (Hetterscheidt et al. 2006).  Note that
 the value of $\sigma_8$, like for example $\sigma_8=0.9$ instead of
 $\sigma_8=0.8$, has virtually no impact on $f_{1/2}$ and, therefore,
 on the measured linear stochastic bias.

 Predicting the power spectra, $P_\kappa$, $P_{\kappa\rm n}$ and
 $P_{\rm n}$, requires a model for the redshift evolution of the 3D
 power spectrum $P_{\rm m}(k,w)$. We use the standard prescription of
 linear structure growth and the Peacock and Dodds (1996) prescription
 for the evolution in the non-linear regime. A more recent and more
 accurate description of the non-linear power spectrum is given by
 Smith et al. (2003). Although Smith et al.  predict in general more
 clustering on smaller scales than Peacock~\&~Dodds, we found in
 a comparison between both methods only little difference for $f_{1/2}$.
 
 It becomes clear from Fig.  \ref{calibrationfig} that particularly
 the interpretation of the bias factor, $b$ (calibration $f_1$),
 depends on $\Omega_{\rm m}$. For the final calibration of the GaBoDS
 measurements we assume as fiducial cosmological model $\Omega_{\rm
   m}=0.3,\,\Omega_\Lambda=0.7,\,\sigma_8=0.8,\,\Gamma=0.21,\,h=0.7$.

 \stressa{The calibration procedure outlined here works because
   correlations between fluctuations seen in $N$ and $M_{\rm ap}$
   stemming from different redshifts quickly vanish with increasing
   mutual redshift difference. This allows us to correct for a
   mismatch in the cosmic volumes seen by $N$ and $M_{\rm ap}$ -- for
   the price of making assumptions about the fiducial cosmology,
   though. In fact, the mismatch is quantified by the factor $1/f_2$
   which is the correlation factor of the (projected) fluctuations
   seen in $N$ and those seen in $M_{\rm ap}$ assuming that galaxies
   perfectly trace mass. If this factor is exactly \mbox{$1/f_2=1$}
   for all scales, we will immediately know that $N$ and $M_{\rm ap}$
   probe exactly the same cosmic volume giving equal weight to all
   radial distances.  For our samples and fiducial cosmology, we find
   \mbox{$1/f_2=0.95,0.95,0.84$} (\mbox{FORE-I} to \mbox{FORE-III})
   indicating that the mismatch is relatively small.}

\subsection{Uncertainties in  the calibration factors}

\stressa{Concerning the statistical errors of the calibrated galaxy
  bias parameters it has to be considered that the redshift
  distributions of the galaxy samples are not exactly known (Sect.
  \ref{pzsect}). Since the calibration just involves a rescaling by
  $f_{1/2}$, any relative error in $f_{1/2}$ results in an equal
  relative error in $b(\theta_{\rm ap})$ and $r(\theta_{\rm ap})$. To
  estimate the relative error in $f_{1/2}$,
  \mbox{$\sigma(f_{1/2})/f_{1/2}$}, we use again the three Jackknife
  samples of the redshift distributions that have already been used in
  Sect. \ref{pzsect}.  Every histogram of photometric redshifts of
  every Jackknife sample is fitted by the template distribution
  \Ref{brainerd}.  Then, for our fiducial cosmology and for every
  galaxy sample, we compute $f_{1/2}$ for each Jackknife
  $p(z)$-template. Thus, for any sample we obtain a triplet of values
  for the bias calibration. Then, as in Sect.  \ref{pzsect}, the
  variance between the different $f_{1/2}$'s is used as estimate of
  the $1\sigma$-error of the calibration parameters. Following this
  procedure, we find that the statistical uncertainty in the
  calibration due to cosmic variance uncertainties in $p(z)$ is
  (\mbox{FORE-I} to \mbox{FORE-III}):
  \mbox{$\sigma(f_1)/f_1=16\%,4\%,3\%$} and
  \mbox{$\sigma(f_2)/f_2=4\%,2\%,8\%$}, respectively. Thus, the
  relative error is roughly
  \mbox{$\sigma(f_{1/2})/f_{1/2}\lesssim7\%$}, except for $f_1$ of
  \mbox{FORE-I} where we have \mbox{$\sigma(f_1)/f_1=16\%$}. We expect
  similar relative errors for other cosmological models.}

 \stressa{Another issue is the impact of uncertainties in the fiducial
   cosmological model on the calibration parameters, which in return
   influences the inferred galaxy bias. Using Fig.
   \ref{calibrationfig} we estimate the relative change in $f_{1/2}$
   with \mbox{$\sim\pm7\%$} ($f_1$) and \mbox{$\sim\pm2\%$} ($f_2$)
   when changing \mbox{$\Omega_{\rm m}=0.3$} by $\pm10\%$. Therefore,
   having $\Omega_{\rm m}$ wrong by about $\pm10\%$ introduces a
   systematic error into the bias parameters which is about $\pm7\%$
   for $b$ and about $\pm2\%$ for $r$.}

\subsection{Redshift and scale resolution}
\label{resolution}

 \begin{figure*}
   \begin{center}
     \psfig{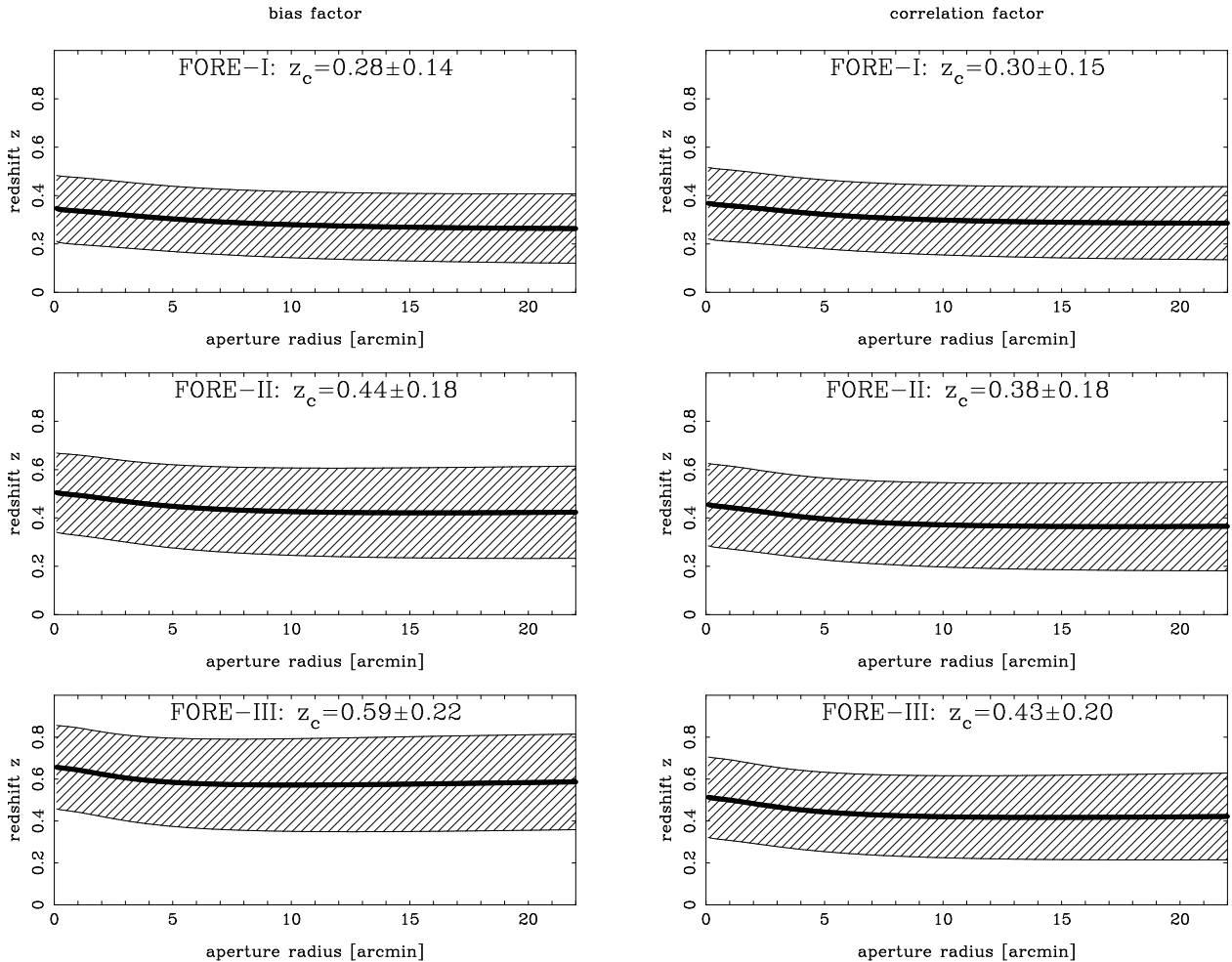}
   \end{center}
   \caption{\label{weight} The bias parameters estimated in this paper
     are averages over some redshift range, plotted here as a function
     of aperture radius; plotted is the mean, $z_{\rm c}$, (solid
     line) and variance, $\sigma_z$, (shaded area) of the weight
     functions $h_{1/3}(z,\theta_{\rm ap})$ for each foreground
     sample. Mean and variance do not change significantly with
     aperture radius, $\theta_{\rm ap}$.  The average of $z_{\rm c}$
     and $\sigma_z$ over all radii, $z_{\rm c}\pm\sigma_z$, is
     indicated in each panel.  \textbf{Left column}: weight function
     for the bias factor, $h_1$.  \textbf{Right column}: weight
     function for the correlation parameter, $h_3$.}
 \end{figure*}

 The bias parameters obtained by Eqs. \Ref{weaklinbias} are in general
 averages of the true bias parameters $b(k,w)$ and $r(k,w)$, Eqs.
 \Ref{truebias}, namely averaged over some scale $k$ and comoving
 distance (redshift) $w$. In other words, the method applied here has
 a limited resolution in redshift and scale. This is due to two
 reasons: 1. the sample of foreground galaxies is usually not peaked
 at one particular redshift but smeared out over some range, and 2.
 lensing is, with varying response, sensitive to the whole matter
 distribution between a source galaxy and the observer. According to
 Hoekstra et al. (2002), the observed weighted averages, $\ave{b^2}$
 and $\ave{r}$, when using the polynomial aperture filter, Eq.
 \Ref{filter}, are approximately
\begin{eqnarray}
  \ave{b^2}(\theta_{\rm ap})&\approx&
  \frac{\int_0^{w_h}\d w\,h_1(w,\theta_{\rm
      ap})\,b^2(\frac{4.25}{\theta_{\rm ap}f_{\rm K}(w)},w)}
  {\int_0^{w_h}\d w\,h_1(w,\theta_{\rm ap})}\;,\\
  \ave{r}(\theta_{\rm ap})&\approx&
  \frac{\int_0^{w_h}\d w\,h_3(w,\theta_{\rm
      ap})\,r(\frac{4.25}{\theta_{\rm ap}f_{\rm K}(w)},w)}
  {\int_0^{w_h}\d w\,h_3(w,\theta_{\rm ap})}\;,
\end{eqnarray}
where the following weight functions have been introduced
\begin{eqnarray}\label{h1}
  h_1(w,\theta_{\rm ap})&=&
  \left(\frac{p_{\rm f}(w)}{f_{\rm K}(w)}\right)^2\,P_{\rm
    filter}(w,\theta_{\rm ap})\;,\\\label{h3}
  h_3(w,\theta_{\rm ap})&=&
  \frac{p_{\rm f}(w)\overline{W}(w)}{[f_{\rm K}(w)]^2}\,P_{\rm
    filter}(w,\theta_{\rm ap})\;,
\end{eqnarray}
with
\begin{eqnarray}\nonumber
&&  P_{\rm filter}(w,\theta_{\rm ap})=\\
&&  2\pi\int_0^\infty\d\ell\,\ell\,P_{\rm m}\left(\frac{\ell}{f_{\rm
        K}(w)\theta_{\rm ap}},w\right)[I(\ell\theta_{\rm ap})]^2\;.
\end{eqnarray}
Note that in general the central redshift and the width of the weights
depends on the aperture radius $\theta_{\rm ap}$.

We calculated the functions $h_1(z,\theta_{\rm ap})$ and
$h_3(z,\theta_{\rm ap})$ for accessible aperture radii for the
redshift distributions in our galaxy samples and for the adopted
fiducial cosmological model. As typical redshift at which $b$ and $r$
are measured we take -- for every aperture radius $\theta_{\rm ap}$ --
the mean, $z_{\rm c}$, of the kernels $h_1$ and $h_3$, and as a
measure for the redshift range over which the bias parameters are
averaged we take the width, $\sigma_z$, of the kernels $h_1$ and $h_3$
(see Fig.  \ref{weight}).  With $z_{\rm c}$ being the typical
redshift, the typical spatial scale corresponds to $k_{\rm
  c}=\frac{4.25}{\theta_{\rm ap}f_{\rm K}(w(z_{\rm c}))}$ (3D Fourier
mode) or $R_{\rm c}=2\pi/k_{\rm c}\approx1.48\,\theta_{\rm ap}f_{\rm
  K}(w(z_{\rm c}))$; the redshift resolution achieved is roughly
$\sigma_z$ which also adds an uncertainty to the effective scale
probed, $\sigma_R$, with a relative error of about $\sigma_R/R_{\rm
  c}=\sigma_z/z_{\rm c}$.

\begin{figure}
  \begin{center}
    \psfig{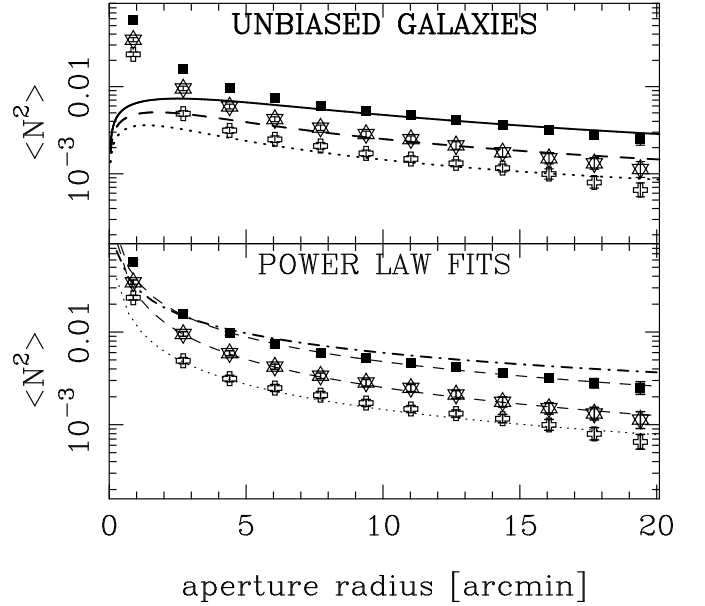}
  \end{center}
  \caption{\label{nsqd}The aperture number count dispersions, as
    measured in GaBoDS, for \mbox{FORE-I} (filled boxes),
    \mbox{FORE-II} (open stars) and \mbox{FORE-III} (open crosses).
    The $1\sigma$ error bars have the size of the data points.
    \textbf{Upper panel}: comparison to $\Lambda\rm CDM$ predictions
    assuming \emph{unbiased galaxies}, upper to lower line:
    \mbox{FORE-I} (solid), \mbox{FORE-II} (dashed) and \mbox{FORE-III}
    (dotted).  \textbf{Lower panel}: power laws give excellent
    descriptions of the measurements. The dotted-dashed line denotes
    $\ave{N^2}$ as measured by Hoekstra et al. (2002)
    ($A_\omega=0.115$, $\delta=0.7$).}
 \end{figure}

\subsection{Combining measurements from different fields}

\paragraph{Bootstrapping I.} The whole data set taken into account for
the analysis consists of $N_{\rm p}=52$ fields. For each field, we
compute the two-point correlation estimates as in Sect.
\ref{usingtwoptcorrelators} and transform them according to the
integrals Eqs.  \Ref{integraltrans1}-\Ref{integraltrans2} to the
$2^{\rm nd}$-order aperture statistics. For every individual field,
$j$, we make $200$ bootstrap samples of the foreground and background
object catalogues to estimate the statistical error of the
measurements 
\begin{equation}
  x_i^{(j)}\!\!\in\!\!\{\ave{N^2(\theta_{{\rm ap},i})},
  \ave{N(\theta_{{\rm ap},i})M_{{\rm
        ap}}(\theta_{{\rm ap},i})}, \ave{M^2_{{\rm
        ap}}(\theta_{{\rm ap},i})}\}\; ,
\end{equation}
for the various aperture radii $\theta_{{\rm ap},i}$. The aperture
moments $x_i^{(j)}$ from all fields are combined to one final result,
$\bar{x}_i$, by the weighted average
\begin{equation}\label{patchaverage}
  \bar{x}_i=
  \frac{\sum_{j=1}^{N_p}x_i^{(j)}\,w_{ij}}
  {\sum_{j=1}^{N_p}\,w_{ij}}
  \; .
\end{equation}
As weight we use the reciprocal bootstrapping variance of $x_i^{(j)}$
in the individual fields, $w_{ij}=1/\sigma^2(x_i^{(j)})$. This
weighting scheme yields the minimum-variance estimate of the average
$\bar{x}_i$.

\paragraph{Bootstrapping II.} In a second bootstrapping stage, we
randomly draw fields from the set of $52$ GaBoDS fields (with putting
back) and calculate a combined signal according to Eq.
\Ref{patchaverage} for this bootstrap sample. We repeat this procedure
$2000$ times. This estimates the PDF of statistical errors in the
\emph{combined signal} including cosmic variance and accounting for
the adopted weighting scheme. In addition to the combined signal of
the aperture moments, we use this technique to also estimate the PDF
of statistical errors of the bias parameters $r(\theta_{\rm ap})$ and
$b(\theta_{\rm ap})$, Eqs. \Ref{weaklinbias}, which are computed from
the weighted average of the aperture statistics. Based on the
bootstrap samples we obtain the median, (asymmetric) $68\%$ confidence
intervals about the median and covariances of the errors for the final
results.

\section{Results}
\label{resultsection}

\subsection{Aperture statistics}

 \begin{figure*}
   \begin{center}
     \psfig{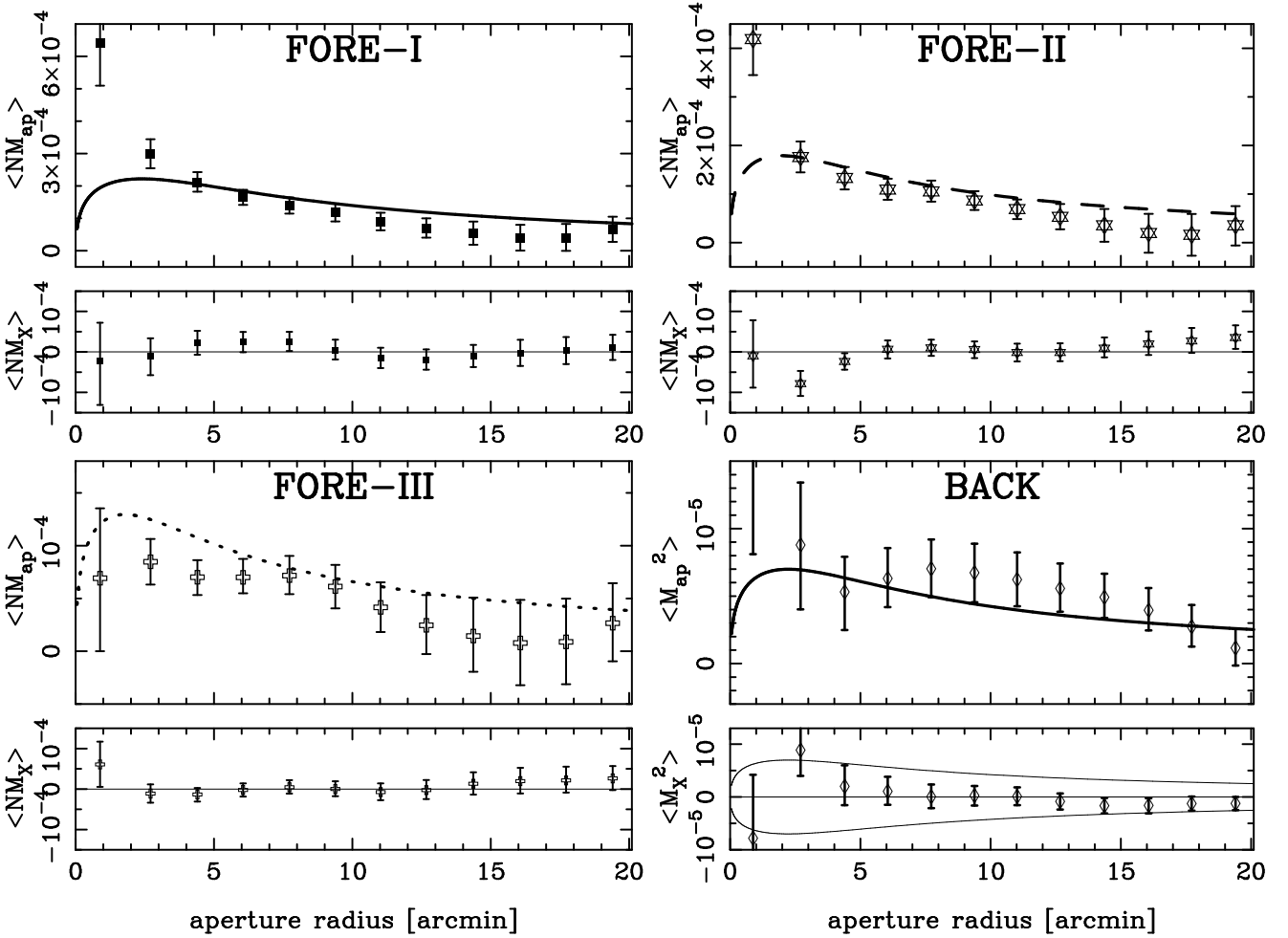}
   \end{center}
   \caption{\label{nmapmapsqd}\textbf{Top row panels and lower left
       panel}: cross-correlation between aperture mass and aperture
     number count for the three different foreground samples
     \mbox{FORE-I} (solid boxes), \mbox{FORE-II} (open stars) and
     \mbox{FORE-III} (open crosses). The panels are subdivided; the
     lower panel shows the B-mode, upper panel is the E-mode of
     $\ave{NM_{\rm ap}}$. The curves are $\Lambda\rm CDM$ predictions
     assuming \emph{unbiased galaxies}.  \textbf{Lower right}:
     aperture mass dispersion, lower and upper panel are B-mode and
     E-mode, respectively. The solid line is a $\Lambda\rm CDM$
     prediction.  The solid lines in the B-mode panel are the E-mode
     prediction with positive and negative sign, which have been
     inserted for comparison.}
 \end{figure*}

 The combined measurements for $\ave{N^2}$, $\ave{NM_{\rm ap}}$ and
 $\ave{M^2_{\rm ap}}$ can be found in Fig. \ref{nsqd} and Fig.
 \ref{nmapmapsqd}.

 \paragraph{Galaxy clustering.} Traditionally, galaxy clustering is
 studied using the angular correlation function $\omega(\theta)$. For
 a comparison of our results for the two-point statistics of galaxy
 clustering with the literature it would be convenient to have
 $\omega(\theta)$ cleaned from the integral constraint. Already from
 early studies on galaxy clustering (e.g.  Davis~\&~Peebles 1983) up
 to recent studies (Norberg et al.  2001; Zehavi et al. 2002) it is
 known that $\omega(\theta)$ is close to a power law over a wide
 range, thus
\begin{equation}
  \omega(\theta)=A_\omega\,\left(\frac{\theta}{1^\prime}\right)^{-\delta}
  \; ,
\end{equation}
where $A_\omega$ and $\delta$ are the clustering amplitude at
$1^\prime$ and the clustering power law index, respectively. Assuming
this power law, we can calculate the aperture number count dispersion
for our polynomial aperture filter:
\begin{equation}\label{nsqdomega}
  \ave{N^2}(\theta)
  =\int_0^2\d x\,x\,\omega(\theta\,x)\,T_+(x)
  =f(\delta)\,A_\omega\,\left(\frac{\theta}{1^\prime}\right)^{-\delta},
\end{equation}
with $f(\delta)$ being defined in \Ref{fdelta}. Therefore, if
$\omega(\theta)$ is a power law, the aperture number count dispersion
is also a power law with the same slope but different amplitude. Since
$\ave{N^2}$ is free of the integral constraint, we can recover
$\omega(\theta)$ from $\ave{N^2}$ by fitting a power law and rescaling
the best-fit amplitude by means of the function \Ref{fdelta}. For
determining the statistical uncertainties of $A_\omega$, however, we
have to take into account that it is not only a function of the
amplitude of $\ave{N^2}$ but also a function of the slope $\delta$.

We performed fits of \Ref{nsqdomega} to the data taking into account
the covariances of $\ave{N^2}$ obtained by bootstrapping, see Fig.
\ref{nsqd}. What is concluded for our foreground samples is summarised
in Table \ref{powfit}.

\begin{table}
  \stressa{
  \caption{\label{powfit}Amplitude and slope of the angular correlation,
    $\omega(\theta)=A_\omega\,(\theta/1^\prime)^{-\delta}$, in our
    foreground galaxy samples as inferred from $\ave{N^2}(\theta_{\rm
      ap})$; $\chi^2/n$ denotes the reduced $\chi^2$ (\mbox{$n=12-2$}) of the
    maximum-likelihood fit.}
\begin{center}
\begin{tabular}{lccc}
  galaxy sample&$A_\omega$&$\delta$&$\chi^2/n$\\\hline\\
  FORE-I&$0.086\pm0.006$&$0.89\pm0.05$&$1.0$\\
  FORE-II&$0.044\pm0.003$&$0.99\pm0.04$&$0.6$\\
  FORE-III&$0.026\pm0.002$&$0.90\pm0.04$&$1.8$\\
\end{tabular}
\end{center}}
\end{table}

The angular correlation of the galaxies in \mbox{FORE-I} -- a sample
roughly comparable to the foreground sample of Hoekstra et al. (2002)
-- has a slope slightly steeper than what is found in the sample of
Hoekstra et al. (there $\delta=0.7$ and $A_\omega=0.115$) and is
smaller in amplitude for aperture radii larger than $\theta_{\rm
  ap}\approx3^\prime$.  This discrepancy in $A_\omega$ and $\delta$ is
not as drastic as it may seem if one takes into account that the
errors of $A_\omega$ and $\delta$ are anti-correlated: a smaller
$A_\omega$ results in a steeper $\delta$. Another issue that may play
a role in this context is the fact that Hoekstra et al. use a
different filter, $R_c$, which is somewhat different from our $R$-band
filter. All in all we think that the measurement of $\omega(\theta)$
for \mbox{FORE-I} is consistent with the measurement of Hoekstra et
al.

Compared to the $\Lambda\rm CDM$ prediction of $\ave{N^2}$ for
unbiased galaxies, which trace the dark matter distribution, our
measurements are clearly different, namely exceeding the dark matter
expectation on scales smaller than \mbox{$\theta_{\rm
    ap}\approx5^\prime$}, and falling slightly below the prediction
for the largest aperture radii. This already suggests a
scale-dependence of the bias factor.

\paragraph{Dark matter clustering.}

\begin{figure}
   \begin{center}
     \psfig{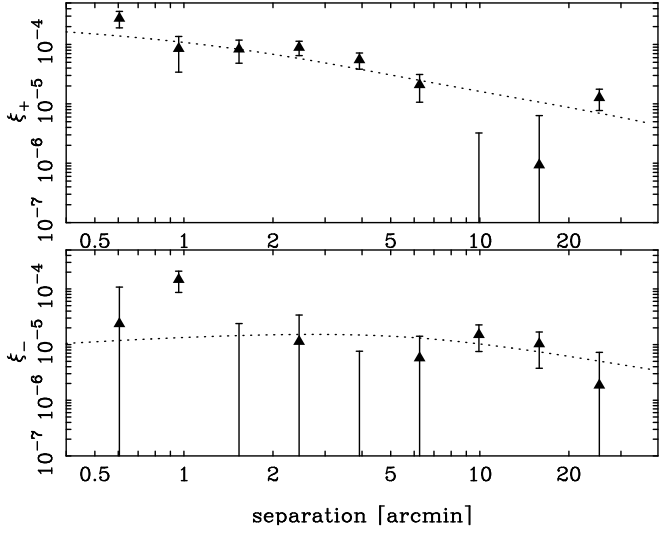}
   \end{center}
   \caption{\label{xipm}The measured two-point cosmic shear
     auto correlation in terms of $\xi_\pm=\ave{\gamma_{\rm
         t}\gamma_{\rm t}}\pm\ave{\gamma_\times\gamma_\times}$. The
     dotted curve is a $\Lambda\rm CDM$ prediction based on $p_{\rm
       b}(w)$ as in the sample BACK.}
 \end{figure}

 \begin{figure}
   \begin{center}
   \psfig{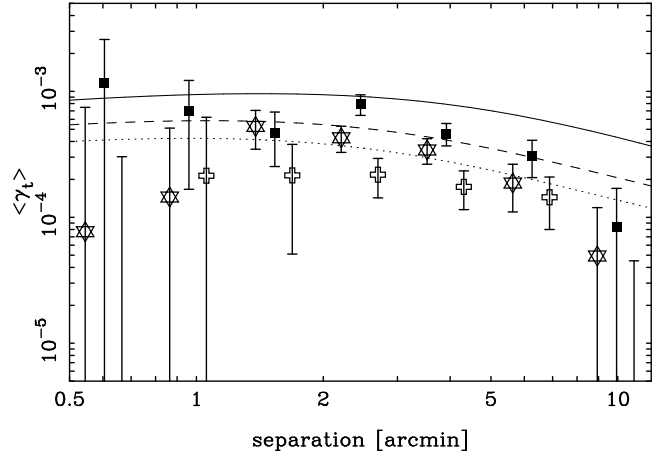}
 \end{center}
 \caption{\label{lambda}Plot of the measured mean tangential shear,
   $\ave{\gamma_{t}}$, about galaxies in our foreground samples
   \mbox{FORE-I} (filled boxes), \mbox{FORE-II} (open stars) and
   \mbox{FORE-III} (open crosses) as a function of angular separation.
   The lines are $\Lambda\rm CDM$ predictions \emph{assuming unbiased
     galaxies}, $p_{\rm f}(w)$ and $p_{\rm b}(w)$ as estimated in
   our samples: for \mbox{FORE-I} (solid), \mbox{FORE-II} (dashed) and
   \mbox{FORE-III} (dotted).}
\end{figure}

The clustering of the total matter content as derived from the
ellipticities of the background galaxies is expressed by the
dispersion of the aperture mass, Fig. \ref{nmapmapsqd}. We calculated
this quantity for a range of different aperture radii from the cosmic
shear two-point correlators, $\xi_\pm$, which are shown in Fig.
\ref{xipm} (rebinned for that plot).

In all figures, the prediction for the adopted fiducial cosmological
model and the estimated redshift distributions in our galaxy samples
is plotted. We conclude that this prediction is in good agreement with
our measurements. Therefore the fiducial cosmology taken for the bias
parameter calibration seems to be reasonable.

Judging from the B-modes, $\ave{M_\times^2}$, in Fig.
\ref{nmapmapsqd}, which serve as an indicator for systematics in the
PSF correction, the PSF correction is ok. Over the whole range of
aperture radii considered the B-modes are consistent with zero, maybe
with a minor exception at about $\theta_{\rm ap}\approx3^\prime$. See
Hetterscheidt et al. (2006) for a detailed discussion on this issue.

\paragraph{Correlation between galaxy and matter distribution.}

The cross-correlation between the $N$-maps and the $M_{\rm ap}$-maps
is plotted in Fig. \ref{nmapmapsqd}. Apart from $\theta_{\rm
  ap}\approx3^\prime$ in \mbox{FORE-II} the B-modes of the signal are
all consistent with zero. The cross-correlation has been worked out on
the basis of the mean tangential shear about galaxies in the
foreground samples.  Results for the galaxy-galaxy lensing signal are
depicted in Fig.  \ref{lambda}.

The data points (E-mode) on intermediate scales are below the
theoretical prediction for $\ave{NM_{\rm ap}}$ based on an unbiased
galaxy population.  This again indicates that either the bias factor
or the correlation parameter or both differ from unity, hinting
towards a population of galaxies that does not perfectly trace the
(dark) matter distribution.

\subsection{Galaxy bias parameters}

 \begin{figure*}
   \begin{center}
     \psfig{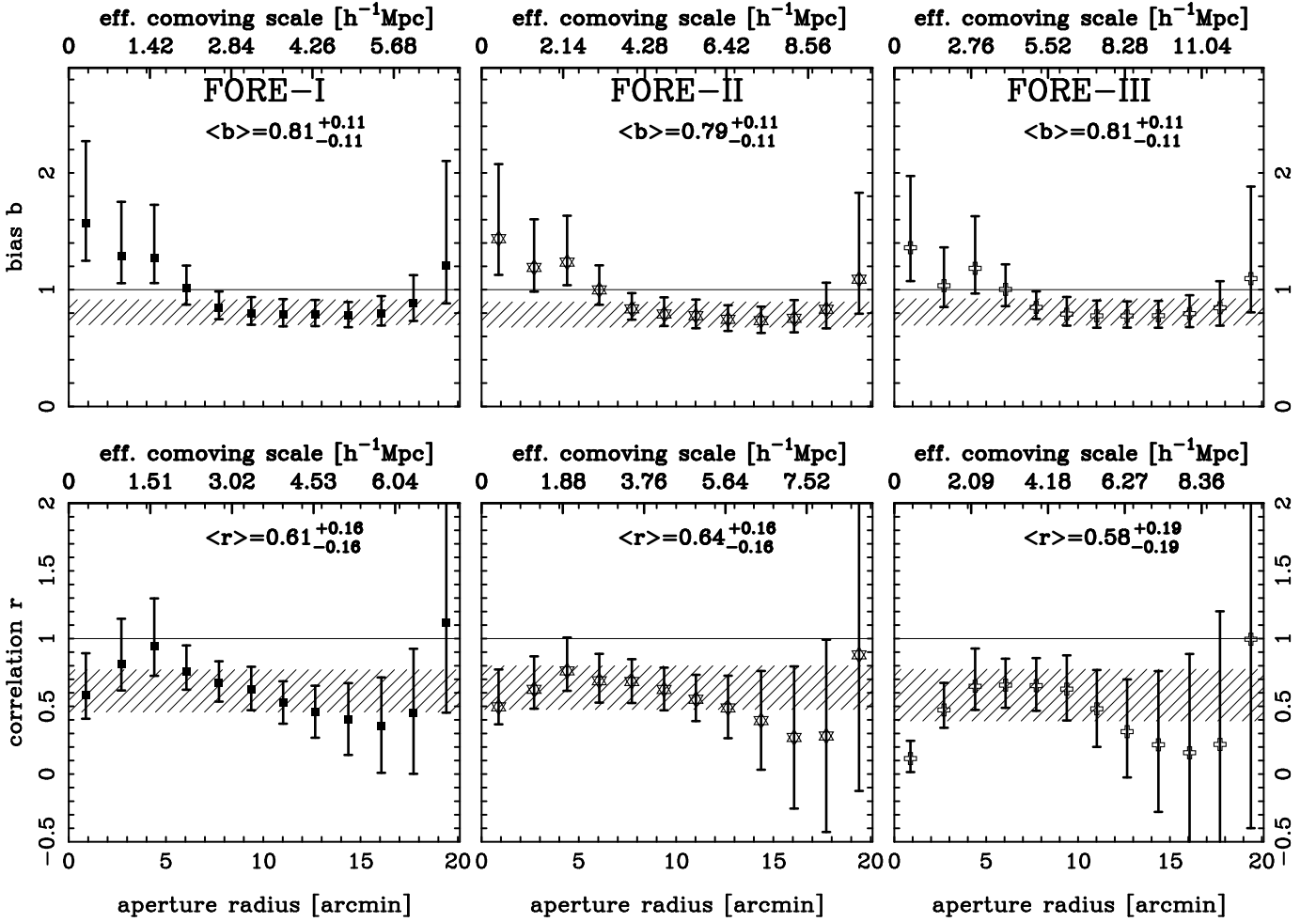}
   \end{center}
   \caption{\label{biasresult} The linear stochastic bias parameters
     of galaxies in the samples \mbox{FORE-I}, \mbox{FORE-II} and
     \mbox{FORE-III} (left to right column); the bias factor, $b$, is
     upper, the correlation parameter, $r$, is in the lower row. The
     parameters have been calibrated assuming $\Omega_{\rm m}=0.3$ and
     $\Omega_\Lambda=0.7$ (see Fig. \ref{calibrationfig}).  The
     effective comoving scale is based on the aperture radius and the
     mean redshift of the weight functions $h_{1/3}$, Fig.
     \ref{weight}.  The bias parameters for a particular aperture
     radius are averages over different physical scales and redshifts
     (Sect.  \ref{resolution}). The shaded area denotes the average
     bias factor or correlation factor over all aperture radii between
     $\theta_{\rm ap}=2^\prime\ldots 19^\prime$; the
     maximum-likelihood of this average and its statistical
     uncertainty are shown in numbers inside the panels.}
 \end{figure*}

 \begin{figure*}
   \begin{center}
     \psfig{file=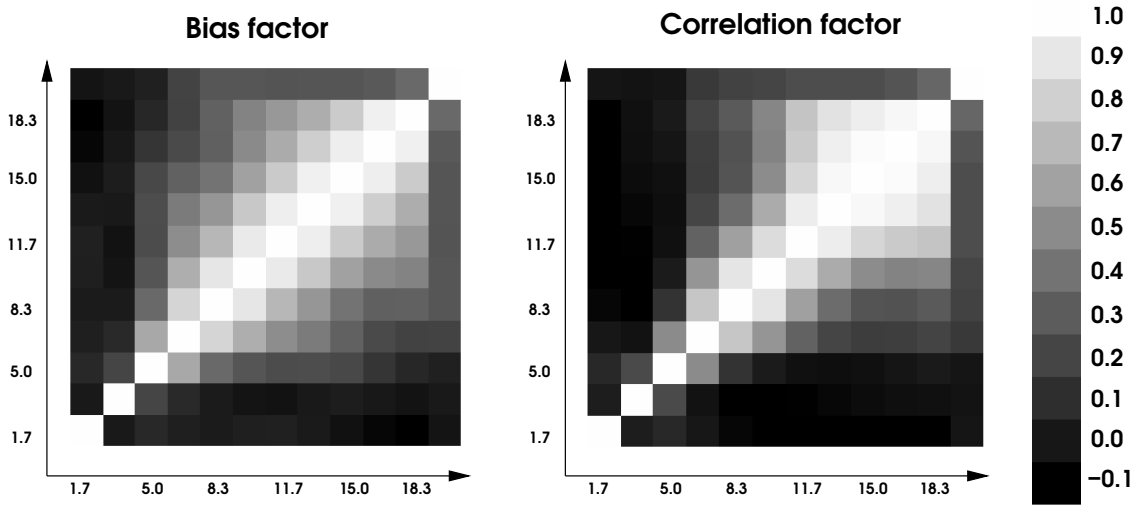,width=150mm,angle=0}
   \end{center}
   \caption{\label{cov} Correlations of the statistical errors of the
     bias factor (left panel) and correlation factor (right panel) of
     galaxy sample \mbox{FORE-II} as inferred from the bootstrap
     samples; the correlation matrices of samples \mbox{FORE-I} and
     \mbox{FORE-III} are virtually identical. The colour of one pixel
     in an intensity map denotes the correlation between the errors
     belonging to the two bins defined by the x- and y-axis; the key
     of the intensity map is on the right side. The numbers attached
     to the axis denote the aperture radii in arcmin corresponding to
     the individual bins in Fig.  \ref{biasresult}.}
 \end{figure*}

 The final result of our work is displayed in Fig. \ref{biasresult}.
 The bias parameters calculated from the aperture statistics, Eqs.
 \Ref{weaklinbias}, have been calibrated, and the aperture radii have
 been converted into a typical physical scale, $R$, based on the mean
 redshift of the range over which the parameters are averaged. As this
 redshift range stretches over about $40\%-50\%$ ($1\sigma$) of the
 mean redshift (see Fig. \ref{weight}), there is a relative
 uncertainty attached to the physical range, $R$, which is of the same
 order; for instance for $R=6\,h^{-1}\rm Mpc$ we have as resolution
 for the effective scale $\sigma_R=3\,h^{-1}\rm Mpc$ (see Sect.
 \ref{resolution}).

 Over the range of (comoving) physical scales investigated, below
 about $R\lesssim10\,h^{-1}\rm Mpc$, the bias factor stays more or
 less constant, rising towards smaller and possibly also larger scales
 with a valley on intermediate scales, where $b$ becomes slightly
 inconsistent with $b=1$ at a $68\%$ confidence level; this implies a
 scale-dependence of the bias factor.  As absolute minimum we obtain
 $b_{\rm min}=0.78\pm0.10,0.74\pm0.10,0.78\pm0.10$ at roughly
 $\theta_{\rm ap}\approx10^\prime$. The position of the minimum is not
 well defined, however, due its width.  In order to get an average
 value for the bias factor, we make a maximum likelihood fit assuming
 a constant bias over the range $2^\prime\le \theta_{\rm ap}\le
 19^\prime$ while taking into account the covariance between the
 errors, as estimated from the bootstrap samples, shown in Fig.
 \ref{cov}.  This fit yields:
 $\bar{b}=0.81\pm0.11,0.79\pm0.10,0.81\pm0.11$ for \mbox{FORE-I},
 \mbox{FORE-II} and \mbox{FORE-III}, respectively.  Therefore, over
 the selected range of scales, galaxies are anti-biased, i.e. less
 clustered than the dark matter.

 The correlation factor, $r$, has a larger relative uncertainty than
 the bias factor, $b$, since it is based on two lensing quantities,
 $\ave{NM_{\rm ap}}$ and $\ave{M_{\rm ap}^2}$, which are generally
 noisier than $\ave{N^2}$. Broadly speaking, the correlation of the
 galaxies to the (dark) matter distribution is relatively high. A
 scale-dependence of the correlation factor is hard to determine due
 to the large uncertainties and the high correlation of neighbouring
 bins; it may be present in the sample \mbox{FORE-I}.  Averaging the
 correlation factor over \mbox{$2^\prime\le \theta_{\rm ap}\le
   19^\prime$} yields
 \mbox{$\bar{r}=0.61\pm0.16,0.64\pm0.16,0.58\pm0.19$} (\mbox{FORE-I}
 to \mbox{FORE-III}) which reflects both the high correlation and the
 unfortunately still large error bars. Obviously, a much larger survey
 area is required to obtain better constraints.  We are going to
 discuss our results in the following section.

\section{Discussion and conclusions}
\label{finalsection}

Observationally, the galaxy-dark matter bias can be probed by means of
various methods (see introduction). Gravitational lensing provides a
promising new method in this respect.  It is special because it allows
for the first time to map the total matter content (mainly dark
matter) with a minimum of assumptions and independent of the galaxy
distribution.  Such a map can be compared to the distribution of
galaxies, or particular types of galaxies, in order to investigate the
galaxy bias.  In particular, correlations between galaxy and dark
matter density become directly visible.  For working out the
galaxy-dark matter bias, older methods rely on assumptions regarding
the growth of dark matter density perturbations, the peculiar
velocities of galaxies and their correlation to the dark matter
density.  Moreover, they often only allow one to measure the bias on
large (linear) scales, \mbox{$\gtrsim 8\,h^{-1}\rm Mpc$}, whereas the
non-linear regime is also accessible with lensing.  However,
gravitational lensing has the disadvantage that it is not equally
sensitive at all redshifts.  The cosmic shear signal is most sensitive
to matter fluctuations roughly half-way between \mbox{$z=0$} and the
mean redshift of the background. This defines a natural best-suited
regime for the method at a redshift of about \mbox{$z\approx0.5$},
often even slightly lower, considering the depth of current galaxy
surveys. It is expected that the most sensitive regime will be shifted
towards higher redshifts by future space-based lensing surveys.
Furthermore, lensing observables are quite noisy so that large survey
areas are required for a good signal-to-noise. Impressively large
surveys with instruments such as the CFHT (CFHT-Legacy-Survey,
CFHTLS), the VST (Kilo-Square-Degree-Survey, KIDS), Pan-STARRS, or
SNAP are either ongoing or about to start within the next years,
providing us with plenty of high signal-to-noise information on dark
matter and galaxy clustering.

In this paper, we employed aperture statistics to quantify the
relation between the dark matter and galaxy density. We tested the
evaluation software against Monte Carlo simulated WFI fields, assuming
an unbiased galaxy population, and found that the software is working
to at least a few percent accuracy (Simon 2005).  The data used is the
GaBoDS with restriction to galaxies brighter than $24\,\rm mag$ in the
$R$-band; this allowed us to estimate the redshift distribution of the
galaxies on the basis of three \mbox{COMBO-17} fields (A901, AXAF/CDFS
and S11) for which photometric redshifts in \mbox{$0\le z\lesssim
1.4$} are available. For all the other fields, only $R$-band
magnitudes can be used to select galaxies.  For this selection, we
defined foreground galaxy samples by choosing galaxies from three
$R$-band magnitude bins that have increasingly fainter median
magnitudes.  The sample \mbox{FORE-I} is comparable to the foreground
selection in Hoekstra et al. (2002) who applied the same technique as
we are using here. By means of the photometric redshifts of the
\mbox{COMBO-17} fields we can translate a GaBoDS R-band magnitude
interval into a redshift distribution.  The fainter the bin, the
broader the redshift distribution, while the mean redshift moves to
larger values.  Therefore, only \mbox{FORE-I} has a rather sharp peak
in redshift, while \mbox{FORE-III} stretches between redshifts of
about \mbox{$z=0.1$} and \mbox{$z\approx 0.9$}.  Hence, \mbox{FORE-II}
and \mbox{FORE-III} are averages over a relatively wide range of
redshifts. In order to get narrower distributions in redshifts with
the aim to reconstruct the redshift evolution of biasing, multi-colour
lensing surveys are required. \stressa{Cosmic variance is the main
uncertainty in the estimated redshift distribution. Based on the
field-to-field variance of the photometric redshift distributions we
estimate that this uncertainty translates into a $1\sigma$-uncertainty
of the bias parameters of $\sim7\%$, except for the bias factor, $b$,
in \mbox{FORE-I} which has $\sim16\%$.}

The B-mode of the aperture statistics $\ave{NM_{\rm ap}}$ and
$\ave{M^2_{\rm ap}}$ are used as an indicator for systematics in the
PSF-corrected shapes of the background galaxies (Fig.
\ref{nmapmapsqd}); they cannot be produced by gravitational lensing
and should therefore be pure noise.  We find that the B-modes are
consistent with zero, maybe leaving a small question mark at
$\theta_{\rm ap}\approx3^\prime$. Note that, at least in principle,
physical effects like intrinsic alignments of the source galaxies can
be a source of B-modes in $\ave{M^2_{\rm ap}}$ (on small scales), so
that vanishing B-modes are not the ultimate indicators of PSF
systematics. For $\ave{NM_{\rm ap}}$, however, the only possible
source of B-modes is a violation of a statistical parity-invariance
(Schneider 2003). Therefore, $\ave{NM_{\rm ap}}$ should always be
B-mode free which is clearly the case in our data.

The fit of a theoretical $\ave{M^2_{\rm ap}}$ constructed from our
fiducial cosmology and redshift distribution of source galaxies to the
measured $\ave{M_{\rm ap}^2}$ is an important test for the calibration
of the bias parameter; $\ave{M^2_{\rm ap}}$ is independent of the
galaxy bias. Our data points are consistent with the fiducial
cosmological model and, therefore, we accept the fiducial cosmological
model and the estimated redshift distributions as sufficiently
accurate for our purposes. For fiducial cosmological models different
from ours (but still flat, \mbox{$\Omega_{\rm m}+\Omega_\Lambda=1$},
with negligible baryon density, \mbox{$\Gamma=\Omega_{\rm m}h$}, and
\mbox{$h=0.7$}, and \mbox{$\sigma_8\propto\Omega_{\rm m}^{-0.56}$})
the calibrated bias parameters in Fig. \ref{biasresult} may be scaled
up or down using Fig. \ref{calibrationfig}.  In a related paper
(Hetterscheidt et al. 2006), we discuss in much more detail issues
concerning the creation of source galaxy catalogues and their data
quality, and we determine constraints on cosmological parameters based
on the GaBoDS data. There it can be seen that this cosmic shear
analysis supports our adopted fiducial cosmology. \stressa{An
  uncertainty in the fiducial cosmology adds an additional uncertainty
  to the galaxy bias calibration and therefore the inferred bias
  parameters. For a realistic relative error of $10\%$ in $\Omega_{\rm
    m}$, we estimate this error to $\sim7\%$ for the bias factor, $b$,
  and to $\sim2\%$ for the correlation factor, $r$. Errors given in
  the following do not include calibration uncertainties.}

The result of the galaxy bias measurement is plotted in Fig.
\ref{biasresult}.  Overall, the galaxy bias factor and the correlation
are close to an unbiased population of galaxies, i.e.  \mbox{$b=1$}
and \mbox{$r=1$}.  A possible scale-dependence is indicated for the
bias factor which rises to $b>1$ on scales below \mbox{$\theta_{\rm
ap}\approx 4^\prime$}, falls below $b=1$ on scales of
\mbox{$\theta_{\rm ap}\approx5^\prime$} and possibly rises again on
larger scales. An aperture radius of $\theta_{\rm ap}=4^\prime$
corresponds to an effective comoving scale of
$R=1.4,2.1,2.8\,h^{-1}\rm Mpc$ (\mbox{FORE-I} to \mbox{FORE-III}) with
a relative uncertainty ($1\sigma$) of about $40\%$. The origin of this
uncertainty is due to the fact that we are actually observing averages
of galaxy bias over some redshift (cosmological time) and scale as
illustrated by Fig.  \ref{weight}; the median redshifts for the bias
are $\bar{z}=0.28,0.44,0.59$, respectively.

The median redshift for the correlation parameters are slightly
different from those of the bias factor, here
$\bar{z}=0.30,0.38,0.43$, again with relative widths ($1\sigma$) of
about $40\%$. Thus, the correlation parameters reflect values typical
for a slightly different, more recent cosmological time. This mismatch
arises if the peak redshift of the lensing efficency,
$\overline{W}(w)$, is displaced with respect to the peak redshift of
the foreground sample, as can be seen by Eq.  \Ref{h3}. An alignment
could be achieved by choosing an appropiate background sample for
every foreground sample which was not possible in our case, because we
did not allow background galaxies fainter than $24\,\rm mag$.

Going back to the observed scale-dependence of the bias factor,
galaxies become anti-biased on intermediate scales; they are less
strongly clustered than the matter. In our data, the minimum value of
the bias factor is determined to be \mbox{$b_{\rm min}\sim0.76$}.
This kind of scale-dependence has also been detected by Pen et al.
(2003) (VIRMOS-DESCART survey) and Hoekstra et al.  (2002)
(VIRMOS-DESCART and RCS) which both rely on weak gravitational lensing
to probe galaxy bias. While Pen et al.  use $I$-band luminosities to
select galaxies, which results in a larger value for the minimum bias
factor but at a similar scale of about $R\approx3\,h^{-1}\rm Mpc$
(\mbox{$k=2\pi/R\approx2h{\rm Mpc}^{-1}$}), the data and sample
selection of Hoekstra et al. is relatively similar to our sample
\mbox{FORE-I}; their value of \mbox{$b_{\rm
    min}=0.71_{-0.04}^{+0.06}$} is in agreement ($1\sigma$) with our
measurement, but the quoted scale of $R\approx1h^{-1}\rm Mpc$ is
different. However, as emphasised before, the position of the minimum
is not well defined in our data. Considering the statistical errors
one has to admit that the position of the bias minimum is not well
determined also in the Pen et al.  analysis (their Fig. 19).  Hence,
there is no contradiction between our data and that of the other
authors.

An anti-bias on the scales considered here and a characteristic
``dip'' in the functional form of the bias factor is in concordance
with recent numerical simulations of dark matter structure formation
(Springel et al. 2005; Weinberg et al. 2004; Guzik~\&~Seljak 2001;
Pearce et al.  2001; Yoshikawa et al.  2001; Somerville et al. 2001;
Jenkins et al.  1998).  The scale-dependence is due to the fact that
the galaxy clustering is a power-law over a wide range of scales,
reflected by $\ave{N^2}$ in Fig.  \ref{nsqd}, while the dark matter
clustering has different shape in CDM simulations and in the
observations suggested by, for instance, $\ave{M_{\rm ap}^2}$ in Fig.
\ref{nmapmapsqd}.

For the linear correlation parameter, we observe as Hoekstra et al.
(2002) and Pen et al. (2003) a high correlation between galaxy and
matter distribution. Averaging the measurement of Hoekstra et al. over
the range \mbox{$2^\prime\le\theta_{\rm ap}\le19^\prime$} yields
roughly \mbox{$r\approx0.8$} which is consistent with our average
($1\sigma$). \stressa{Our observed correlations between fluctuations
  in the galaxy number and mass density appear to be a bit lower,
  though (Hoekstra, private communication). This could hint to an
  hitherto undiscovered systematic effect in our data.  However, it
  should be kept in mind that the statistical errors in $r$ are highly
  correlated and quite large so that this slightly lower value of $r$
  may be just a statistical fluke.}  The clear scale-dependence of the
correlation parameter observed by Hoekstra et al. is not visible in
our data, because this feature probably gets lost within the
statistical uncertainties.

The figures for the correlation parameter -- $r$ is smaller than unity
with $68\%$ confidence -- show that the galaxies are either
stochastically or non-linearly biased, or a mixture of both.  To
understand what is meant here, imagine that $\delta_{\rm g}$ -- the
galaxy density contrast -- and $\delta_{\rm m}$ -- the dark matter
density contrast (both smoothed) -- are quite generally related by
\begin{equation}
  \delta_{\rm g}=f(\delta_{\rm m})+\epsilon(\delta_{\rm m})\;,
\end{equation}
where $f$ is some function and $\epsilon$ a random variable (noise),
both satisfying
\begin{equation}
  \ave{f(\delta_{\rm m})}=\ave{\epsilon(\delta_{\rm m})}=\ave{\epsilon(\delta_{\rm
      m})f(\delta_{\rm m})}=0
\end{equation}
owing to the definition of the density contrast and noise ($\epsilon$
is statistically independent of $f$). In the case of $f$ being linear
and $\epsilon=0$ we have a linear and deterministic relation between
$\delta_{\rm g}$ and $\delta_{\rm m}$, whereas $\epsilon\ne 0$
introduces a stochasticity between the density fields. The latter case
is called stochastic bias. If $\delta_{\rm g}$ and $\delta_{\rm m}$
were Gaussian random variables, $f$ would be a linear function. A
non-linear function $f$ yields what is commonly called a non-linear
bias. The degeneracy between non-linearity and stochasticity arises
because a decorrelation -- indicated in the linear stochastic bias
scheme by $r\!\!<\!\!1$ -- can be generated by both a non-linear $f$
and a stochastic component $\epsilon$:
\begin{equation}
  r = \frac{\ave{\delta_{\rm g}\delta_{\rm m}}}{\sqrt{\ave{\delta_{\rm
          g}^2}\ave{\delta_{\rm m}^2}}}=
  \frac{\ave{\delta_{\rm m}f(\delta_{\rm m})}}{\sqrt{\ave{\delta_{\rm
          m}^2}(\ave{[f(\delta_{\rm m})]^2}+\ave{[\epsilon(\delta_{\rm
          m})]^2})}}\;.
\end{equation}

Discriminating between these two cases requires the additional
measurement of the non-linear stochastic bias parameter (Yoshikawa et
al. 2001; Dekel~\&~Lahav 1999) or equivalent quantities. So far, these
parameters have only been measured for the relative bias between
populations of galaxies (Wild et al. 2005). Therefore, it is unknown
from observations whether this decorrelation between galaxies and dark
matter is mainly due to a random scatter or a non-linear relation. To
resolve this, using approaches similar to ours where statistical
moments of the joint PDF of matter and galaxies are measured, one
needs to invoke higher-order statistics. As the currently ongoing
research is working on the three-point statistics of the aperture mass
and the aperture number count (e.g.  Schneider~\&~Watts 2005;
Schneider, Kilbinger~\&~Lombardi 2005; Jarvis et al.  2004;
Schneider~\&~Lombardi 2003) we can expect to be capable of such a task
quite soon.

Within the uncertainties of our measurement we do not see a difference
in the biasing parameters between the three foreground bins.  As the
three different foreground bins represent different median redshifts
of the galaxies, we conclude that on the scales considered the
redshift dependence of the (averaged) linear bias for
\mbox{$0.3\lesssim z\lesssim0.7$} has to be smaller than about $\Delta
b\lesssim0.2$ and $\Delta r\lesssim 0.4$ ($1\sigma$) as crudely
estimated from the $1\sigma$-errors of the average bias and
correlation parameter; any larger bias evolution should have been
detectable despite the relatively large error bars.  These figures are
no serious constraints to cosmological models for the bias evolution
because all different numerical and analytic models predict evolution
rates well below these limits in the redshift range covered here (cf.
Magliocchetti et al. 2000).  In a recent paper, Marinoni et al. (2005)
measured the non-linear biasing function (Dekel~\&~Lahav 1999) in the
VIMOS-VLT Deep Survey between \mbox{$0.7\lesssim z\lesssim1.5$} and
found that the bias evolution is marginal below $z\sim0.8$ and becomes
more pronounced beyond that redshift.  Empirically, they found the
redshift dependence of the bias factor on a scale of $8\,h^{-1}\rm
Mpc$ being described by \mbox{$b(z)=1+(0.03\pm0.01)(1+z)^{3.3\pm0.6}$}
which means a change of $b$ of roughly $5\%$ between $0.3\le z\le
0.7$. This figure is in qualitative agreement with our observation.

The bias parameters, no matter whether linear stochastic or non-linear
stochastic bias, are just conveniently defined quantities for a
comparison of random fields. They bear no obvious relation to the
physics of galaxies. In the end, these measurements will
need to be interpreted in terms of physical quantities like the halo
occupation distribution (Berlind, Weinberg et al.  2003;
Berlind~\&~Weinberg 2002; Peacock~\&~Smith 2000) in order to learn
more about the evolution and formation of galaxies in the environment
of their parent dark matter haloes.


\begin{acknowledgements}
  This work was supported by the Deutsche Forschungsgemeinschaft (DFG)
  under the Graduiertenkolleg 787, the projects ER 327/2--1, SCHN
  342/6--1 and SCHN 342/7--1. Further support was received from the
  Ministry for Science and Education (BMBF) through DESY under the
  project 05AV5PDA/3.  We also acknowledge the support given by
  ASTROVIRTEL, a project funded by the European Commission under FP5
  Contract No.  HPRI-CT-1999-00081. Christian Wolf was supported by a
  PPARC Advanced Fellowship.
\end{acknowledgements}

\appendix

\section{Angular clustering and the aperture number count dispersion}

We will briefly demonstrate in this section that the integral transformation
\Ref{integraltrans2} applied to a offset power law
\begin{equation}\label{powlaw}
  \omega(\theta)=A_\omega\,\theta^{-\delta}-C\;,
\end{equation}
where $C$ is a constant (the integral constraint), results in a power
law aperture number count dispersion $\ave{N^2}$.

For the polynomial aperture filter $u$ used here, Eq. \Ref{filter},
one obtains for the transformation kernel $T_+$ an analytical
expression that can be found in Schneider et al. (2002).  Using this
kernel and Eq. \Ref{powlaw} for $\omega(\theta)$ into
\Ref{integraltrans2} yields
\begin{equation}
  \ave{N^2(\theta_{\rm ap})}=A_\omega\,f(\delta)\,\theta_{\rm ap}^{-\delta}\;,
\end{equation}
where the function $f(\delta)$ has been obtained by
\texttt{Mathematica}\footnote{Wolfram Research:
  \texttt{http://www.wolfram.com}.},
\begin{eqnarray}\label{fdelta}
&&
f(\delta)\equiv\frac{1}{25} {2^{3+\delta}}\times\\&&\nonumber
\Bigg(\frac{30}{2+\delta}-\frac{900}{4+\delta}+\frac{15 \,\Gamma(\frac{3+\delta}{2})}{{\sqrt{\pi }}
\,\Gamma(3+\frac{\delta}{2})}-\\\nonumber&&\frac{30 \bigg((2+\delta) {\sqrt{\pi }}-\frac{2 \,\Gamma(\frac{3+\delta}{2})}{\,\Gamma(1+\frac{\delta}{2})}\bigg)}{{{(2+\delta)}^2}
{\sqrt{\pi }}}+\frac{1508 \,\Gamma(\frac{7+\delta}{2})}{{\sqrt{\pi }} \,\Gamma(5+\frac{\delta}{2})}+\\&&\nonumber
\frac{1160 \,\Gamma(\frac{5+\delta}{2})}{{\sqrt{\pi }} \,\Gamma(4+\frac{\delta}{2})}+\frac{900 \bigg((4+\delta) {\sqrt{\pi }}-\frac{2 \,\Gamma(\frac{5+\delta}{2})}{\,\Gamma(2+\frac{\delta}{2})}\bigg)}{{{(4+\delta)}^2}
{\sqrt{\pi }}}-\\&&\nonumber\frac{1056
\,\Gamma(\frac{9+\delta}{2})}{{\sqrt{\pi }} \,\Gamma(6+\frac{\delta}{2})}-\frac{288 \,\Gamma(\frac{11+\delta}{2})}{{\sqrt{\pi
}} \,\Gamma(7+\frac{\delta}{2})}\Bigg)
\; . 
\end{eqnarray}
Therefore, $\ave{N^2}$ is insensitive to the offset in
$\omega(\theta)$ and is a power law with the same slope as
$\omega(\theta)$.  In the regime \mbox{$\delta\in[0.2,1.6]$}, the
somewhat bulky function $f(\delta)$ can, within a few percent
accuracy, be approximated by
\begin{equation}
  f(\delta)\approx
  0.0051\,\delta^{11.55}+0.2769\,\delta^{3.95}+0.2838\,\delta^{1.25}\;,
\end{equation}
 which covers the commonly observed range of values for the power law
index.



\begin{thebibliography}{99}
\bibitem{BDS98} Baker, J.E., Davis, M., Strauss, M. A., Lahav, O.,
  Santiago, B.X., 1998, ApJ, 508, 6
\bibitem{BBKS86} Bardeen, J.M., Bond, J.R., Kaiser, N., Szalay, A.S.,
  1986, ApJ, 304, 15
\bibitem{BS01} Bartelmann, M., Schneider P., 2001, Phys. Rev., 340,
  291
\bibitem{BMdC96} Benoist, C., Maurogordato, S., da Costa, L.N., et al.,
  1996, ApJ, 472, 452
\bibitem{BW02} Berlind, A.A., Weinberg, D.H., 2002, ApJ, 575, 587

\bibitem{BWB03} Berlind, A.A., Weinberg, D.H., Benson, A.J, \emph{et al.},
  2003, ApJ, 593, 1
\bibitem{BA96} Bertin, E., Arnouts, S., 1996, A\&A, 117, 393
\bibitem{b91} Blandford, R.D., Saust, A.B., Brainerd, T.G. Villumsen,
  J.V., 1991, MNRAS, 251, 600
\bibitem{B00} Blanton, M.R., 2000, ApJ, 544, 63
\bibitem{BBS96} Brainerd, T.G., Blandford, R.D., Smail, I., 1996, Apj,
  466, 623
\bibitem{} Coe, D., Benitez, N., Sanchez, S.F., et al., 2006, accepted
  by AJ, \texttt{astro-ph/0605262}
\bibitem{C01} Colless, M.M., et al. (the 2dFGRS team), 2001, MNRAS,
  328, 1039
\bibitem{} Conway, E., Maddox, S., Wild, V., et al., 2005, MNRAS, 356,
  456
\bibitem{DG76} Davis, M., Geller, M.J., 1976, ApJ, 208, 13
\bibitem{DP83} Davis, M., Peebles, P.J.E., 1983, ApJ, 267, 465
\bibitem{DBY93} Dekel, A., Bertschinger, E., Yahil, A., Strauss, M.A.,
  Davis, M., Huchra, J.P., 1993, ApJ, 412, 1
\bibitem{DL99} Dekel, A., Lahav, O., 1999, ApJ, 520, 24
\bibitem{EKS90} Efstathiou, G., Kaiser, N., Saunders, W., et al., 1990,
  MNRAS, 247, 10
\bibitem{EBW92} Efstathiou, G., Bond, J.R., White, S.D.M., 1992, MNRAS,
  258, 1
\bibitem{EB99} Efstathiou, G., Bond, J.R., 1999, MNRAS, 304, 75
\bibitem{EvWB01} Erben, T., van Waerbeke, L., Bertin, E., Mellier, Y.,
  Schneider, P., 2001, A\&A, 366, 717
\bibitem{ESD05} Erben, T., Schirmer, M., Dietrich, J.P., et al., 2005,
  AN, 326, 432
\bibitem{F00} Fischer, P., McKay, T.A., Sheldon, E., et al., 2000, AJ,
  120, 1198
\bibitem{F96} Fry, J.N., 1996, ApJ, 461, L65
\bibitem{GF94} Gazta\'{n}aga, E.,Frieman, J.A., 1994, ApJ, 437, 13
\bibitem{GY01} Gladders, M.D., Yee, H.K.C., \texttt{astro-ph/0011073}
\bibitem{GP77} Groth, E.J., Peebles, P.J.E., 1977, ApJ, 217, 385
\bibitem{GS01} Guzik, J., Seljak, U., 2001, MNRAS, 321, 439
\bibitem{HS06} Hetterscheidt, M., Simon, P., Erben, T., et al., 2006,
  submitted to A\&A, \texttt{astro-ph/0606571}
\bibitem{HBH04} Heymans, C., Brown, M., Heavens, A., et al., 2004,
  MNRAS, 347, 895
\bibitem{H05} Heymans, C., van Waerbeke, L., Bacon, D., et al., 2006,
  MNRAS, 368, 1323
\bibitem{HMS04} Hirata, C.M., Mandelbaum, R., Seljak, U., et al.,
  2004, MNRAS, 353, 529
\bibitem{HYG01} Hoekstra, H., Yee, H.K.C., Gladders, M.D.,
  2001, ApJ, 558, 11
\bibitem{HvWGM02} Hoekstra, H., van Waerbeke, L., Gladders, M.D., Mellier,
  Y., Yee, H.K.C., 2002, ApJ, 577, 604
\bibitem{HFK03} Hoekstra, H., Franx, M., Kuijken, K., Carlberg, R.G.,
  Yee, H.K.C., 2003, MNRAS, 340, 609
\bibitem{HGD98} Hudson, M.J., Gwyn, S.D.J., Dahle, H., Kaiser, N.,
  1998, ApJ, 503, 531
\bibitem{JBJ04} Jarvis, M., Bernstein, G., Jain, B., 2004, MNRAS, 352,
  338
\bibitem{JFP98} Jenkins, A., Frenk, C.S., Pearce, F.R., et al., 1998,
  ApJ, 499, 20
\bibitem{K87} Kaiser, N., 1987, MNRAS, 227, 1
\bibitem[1992]{kai92} Kaiser, N., 1992, ApJ, 388, 272
\bibitem{KSB95} Kaiser, N., Squires, G., Broadhurst, T., 1995, ApJ,
  449, 460
\bibitem{KSN04} Kayo, I., Suto, Y., Nichol, R.C., et al., 2004, PASJ,
  56, 415
\bibitem{KRE05} Kleinheinrich, M, Rix, H.-W., Erben, T., et al., 2005,
  A\&A, 439, 513
\bibitem{LC93} Lacey, C., Cole, S., 1993, MNRAS, 262, 627
\bibitem{LBP02} Lahav, O., Bridle, S.L., Percival, W.J., et al., 2002,
  MNRAS, 333, 961
\bibitem{LS93} Landy, S.D., Szalay, A.S., 1993, 412, 64
\bibitem{LME95} Loveday, J., Maddox, S.J., Efstathiou, G., Peterson,
  B.A., 1995, ApJ, 442, 457
\bibitem{LEM96} Loveday, J., Efstathiou, G., Maddox, S.J., Peterson,
  B.A., 1996, ApJ, 468, 1
\bibitem{MES90} Maddox, S.J., Efstathiou, G., Sutherland, W.J., Loveday,
  J., 1990, MNRAS, 242, 43
\bibitem{MHL03} Madgwick, D.S., Hawkins, E., Lahav, O., et al., 2003,
  MNRAS, 344, 847
\bibitem{MBML00} Magliocchetti, M., Bagla, J.S., Maddox, S.J., Lahav,
  O., 2000, MNRAS, 314, 546
\bibitem{} Marinoni, C., Le F\`{e}vre, O., Meneux, B., et al., 2005, A\&A,
  442, 801
\bibitem{MSR01} McKay, T.A., Sheldon, E.S., Racusin, J., et al., 2001,
  submitted to ApJ, \texttt{astro-ph/0108013}
\bibitem[1991]{Mir91} Miralda-Escud\'e, J., 1991, ApJ, 380, 1
\bibitem{MW96} Mo, H.J., White, S.D.M., 1996, MNRAS, 282, 347
\bibitem{MCG01} Moore, A.W., et al., \emph{Fast Algorithms and efficient
  statistics: N-Point Correlation Functions}, misk.conf. 71
\bibitem{N01} Norberg, P., Baugh, C.M., Hawkins, E., et al., 2001,
  MNRAS, 328, 64
\bibitem{NBH02} Norberg, P., Baugh, C.M., Hawkins, E., et al., 2002,
  MNRAS, 332, 827
\bibitem{PD96} Peacock, J.A., Dodds, S.J., 1996, MNRAS, 280,
  L19
\bibitem{PS00} Peacock, J.A., Smith, R.E., 2000, MNRAS, 318, 1144\bibitem{PJF01} Pearce, F.R, Jenkins, A., Frenk, C.S., White, S.D.M.,
  et al., 2001, MNRAS, 326, 649
\bibitem{PEN98} Pen, U.-L., 1998, ApJ, 504, 601
\bibitem{PLvW03} Pen, U.-L., Lun, T., van Waerbeke, L., Mellier, Y.,
  2003, MNRAS, 346, 994
\bibitem{PZ03} Pen, U.-L., Zhang, L.L., 2005, New. Astron., 10, 569 
\bibitem{P01} Pierfederici, F., 2001, Proceedings SPIE 4477, 246
\bibitem{} Schirmer, M., 2004, Doctoral thesis, ``\emph{Weak
    gravitational lensing: Detection of mass concentrations in wide
    field imaging data}'', University of Bonn, Germany
\bibitem{S96} Schneider, P., 1996, MNRAS, 283, 837
\bibitem{S98} Schneider, P., 1998, ApJ, 498, 43
\bibitem{SvWM02} Schneider, P., van Waerbeke, L., Mellier, Y.,
  2002, A\&A 389, 729
\bibitem{} Schneider, P., Lombardi, M., 2003, A\&A, 397, 809
\bibitem{} Schneider, P., 2003, A\&A, 408, 829
\bibitem{S04} Schneider, P., Kochanek, C., Wambsganss, J., 2006,
  Saas-Fee Advanced Course 33, ``\emph{Gravitational Lensing: Strong,
    Weak and Micro}'', Swiss Society for Astronomy and Astrophysics,
  ISBN: 3-540-40409-X
\bibitem{SKL04} Schneider, P., Kilbinger, M., Lombardi, M., 2005,
  A\&A, 431, 9
\bibitem{} Schneider, P., Watts, P., 2005, A\&A, 432, 738
\bibitem{SS97} Seitz, C., Schneider, P., 1995, A\&A, 318, 687
\bibitem{SMM04} Seljak, U., Makarov, A., Mandelbaum, R., et al., 2005,
  Phys.Rev. D, 71, 043511
\bibitem{SLO96} Shectman, S.A., Landy, S.D., Oemler, A., et al., 1996,
  ApJ, 470, 172
\bibitem{SJF04} Sheldon, E.S., Johnston, D.E., Frieman, J.A., et al.,
  2004, AJ, 127, 2544
\bibitem{SBD00} Sigad, Y., Branchini, E., Dekel, A., 2000, ApJ, 540, 62
\bibitem{SED98} Sigad, Y., Eldar, A., Dekel, A., Strauss, M., Yahil
  A., 1998, ApJ, 495, 516
\bibitem{S05} Simon, P., Doctorial thesis, ``\emph{Weak Gravitational
    Lensing and Galaxy Bias}'', 2005, University of Bonn, Germany
\bibitem{S06} Simon, P., 2006, submitted to A\&A, \texttt{astro-ph/0609165}
\bibitem{SPJ03} Smith, R.E., Peacock, J.A., et al., 2003, MNRAS, 341,
  1311
\bibitem{SLS01} Somerville, R.S., Lemson, G., Sigad, Y., Dekel, A.,
  Kauffmann, G., White, S.D.M., 2001, MNRAS, 320, 289
\bibitem{S05} Springel, V., White, S.D.M., Jenkins, A., et al., 2005,
  Nature, 435, 629
\bibitem{TP98} Tegmark, M., Peebles, P.J.E., 1998, ApJ, 500, L79
\bibitem{TB99} Tegmark, M., Bromley, B., 1999, ApJ, 518, L69
\bibitem{TSB04} Tegmark, M., Strauss, M.A., Blanton, M.R., et al.,
  2004, PhysRevD, 69, 103501
\bibitem{TOK97} Tucker, D.L., Oemler, J.A., Kirshner, R.P., et al.,
  1997, MNRAS, 285, 1335
\bibitem{vW98} van Waerbeke, L., 1998, A\&A, 334, 1
\bibitem{vWMR01} van Waerbeke, L., Mellier, Y., Radovich, M., et
  al., 2001, A\&A, 374, 757
\bibitem{vWM03} van Waerbeke, L., Mellier, Y., 2003, Gravitational
  Lensing by Large Scale Structures: A Review, lecture given at
  Aussois Winter School, 01/2003, \texttt{astro-ph/0305089}
\bibitem{vW06} van Waerbeke, L., White, M., Hoekstra, H., Heymans, C.,
  2006, \texttt{astro-ph/0603696}
\bibitem{VHP02} Verde, L., Heavens, A.F., Percival, W.J., 2002, MNRAS,
  335, 432
\bibitem{WDK04} Weinberg, D.H., Dave, R., Katz, N., Hernquist, L.,
  2004, ApJ, 60, 1
\bibitem{WR78} White, S.D.M., Rees, M.J., 1978, MNRAS, 183, 341
\bibitem{WF91} White, S.D.M., Frenk, C.S., 1991, ApJ, 379, 52
\bibitem{WEF93} White, S.D.M., Efstathiou, G., Frenk, C.S., 1993,
  MNRAS, 1023
\bibitem{WPL04} Wild, V., Peacock, J.A., Lahav, O., et al., 2005,
  MNRAS, 356, 247
\bibitem{WKL01} Wilson, G., Kaiser, N., Luppino, G.A., 2001, ApJ, 556,
  601
\bibitem{WMK04} Wolf, C., Meisenheimer, K., Kleinheinrich, M., et al.,
  2004, A\&A, 421, 913
\bibitem{Y00} York, D.G., et al. (the SDSS team), AJ, 120, 1579
\bibitem{YTJ01} Yoshikawa, K., Taruya, A., Jing, Y.P., Suto, Y., 2001,
  ApJ, 558, 520
\bibitem{ZBF02} Zehavi, I., Blanton, M.R., Frieman, J.A., et al.,
  2002, ApJ, 571, 172  

\end{thebibliography}
\end{document}